\documentclass[10pt]{report}  

\usepackage[intoc]{nomencl}

\textfloatsep = 0.4in \addtocontents{toc}{\vspace{0.4in} \hfill
Page\endgraf} \addtocontents{lof}{\vspace{0.2in} \hspace{0.13in} \
Figure\hfill Page\endgraf} \addtocontents{lot}{\vspace{0.2in}
\hspace{0.13in} \ Table\hfill Page\endgraf}

\usepackage{textcomp}
\usepackage{array}
\usepackage{listings}
\usepackage{setspace}
\usepackage{mathptmx}
\usepackage[table, svgnames]{xcolor}
\usepackage{colortbl}
\usepackage{graphicx}
\usepackage{amssymb, amsmath}
\usepackage{subfig}
\usepackage{epsfig}
\usepackage{times}
\usepackage{float}
\usepackage{rotating}
\usepackage{makeidx}
\usepackage{url}
\usepackage{multirow}
\usepackage{booktabs}
\usepackage{tabularx}

\usepackage[subfigure, titles]{tocloft}
\usepackage{acronym}
\usepackage{datetime}

\usepackage{algorithm}
\usepackage{algorithmic}

\makenomenclature

\graphicspath{{Figures/}}
\DeclareGraphicsExtensions{.pdf,.jpeg,.png,.PNG, .eps, .tiff}

\usepackage{makecell}
\usepackage{titletoc}
\usepackage{sfchap}
\usepackage{sfsection}
\usepackage[authoryear]{natbib}
\usepackage{appendix}
\usepackage[nottoc]{tocbibind}
\usepackage{hyperref}
\hypersetup{
	pdftitle={Towards Single Slot Finality: Evaluating Consensus Mechanisms and Methods for Faster Ethereum Finality},
	pdfauthor={Lincoln Murr},
	bookmarksnumbered, 
	pdfstartview={FitH},
	pdfborder={0 0 0},
	plainpages=false
}%
\usepackage[all]{hypcap}

\begin{document}

\pagenumbering{alph}

\begin{titlepage}
\thispagestyle{empty}\enlargethispage{\the\footskip}%
\begin{center}
    \hspace{1em}\vspace{6em} \\ 
	{\setstretch{1.66} \MakeUppercase{Towards Single Slot Finality: Evaluating Consensus Mechanisms and Methods for Faster Ethereum Finality}\par }%
	\vskip.5in
	By
	\vskip .4in
	{Lincoln Dietz Murr}
	\vskip .4in
	
	\begin{doublespace}
	Thesis\\
		Submitted to the Faculty of the \\
		Graduate School of Vanderbilt University \\
		in partial fulfillment of the requirements \\
		for the degree of \\ [.2in]
	\end{doublespace}
	
	MASTER OF SCIENCE \\[.1in]
	in \\[.2in]
	Computer Science \\[.2in]
	May 10, 2024 \\[.2in]
	
	Nashville, Tennessee
	\vskip .5in
\end{center}



\begin{center}
\begin{doublespace}
Approved:\\ [.2in]
Dana Zhang, Ph.D. \\  [.1in]
Douglas C. Schmidt, Ph.D. \\ [.1in]
\end{doublespace}
\end{center}
\end{titlepage}

\doublespacing
\pagenumbering{roman} \setcounter{page}{2}

\singlespacing

\chapter*{}
\hspace{1em}\vspace{18em} \\ 
\begin{center}
    Copyright {\textcopyright} 2024 Lincoln Dietz Murr \\
    All Rights Reserved     
\end{center}


\chapter*{ACKNOWLEDGMENTS}
\vspace{7mm}

Thank you to the Ethereum Foundation for supporting this work through the Ethereum Protocol Fellowship.

\singlespacing
\tableofcontents

\begingroup
\setlength{\parskip}{1\baselineskip}
\endgroup

\normalsize
\doublespacing
\pagenumbering{arabic}
\setcounter{page}{1}

\chapter{Introduction}
Distributed ledger technology, particularly blockchain, has solved the complex problem of creating a mechanism by which entities can coordinate and communicate in a trustless and permissionless manner. The key innovation was the integration of distributed consensus mechanisms with economic incentives for validators, creating a system in which validators are incentivized to act honestly and in the network's best interest.

Ethereum has emerged as one of the most widely adopted blockchain platforms, supporting a global computing infrastructure through a decentralized network of nodes. At its core, Ethereum relies on a novel consensus protocol called Gasper, which combines the LMD-GHOST fork choice rule with the Casper Friendly Finality Gadget (FFG) overlay for the probabilistic finality of transactions. Combined with a proof-of-stake collateralization requirement for validators, these rules have created a robust system that satisfies both security and liveness. 

While this construction has proven effective, limitations exist in guaranteeing fast transaction finality. The current model requires a confirmation time of 64 to 95 blocks, approximately 15 minutes, before Casper finalizes a block. This delay impacts user experience and exposes the network to short-term chain reorganizations, potentially allowing transaction censorship or manipulation by validators with no severe penalties. With the recent emphasis on a rollup-centric roadmap as the path for scaling Ethereum, faster finality would also allow for cross-layer and inter-rollup communication to take place significantly faster, as rollups would no longer have to wait an extended period for finality before approving a deposit. Furthermore, other blockchains and their respective consensus mechanisms are capable of faster finality, in some cases instantaneous finality, because of innovations over prior solutions and tradeoffs in liveness or decentralization. Achieving single slot finality (SSF), where transactions are proposed and finalized within the same slot, has emerged as a path to expand the capabilities of Ethereum.

The organization of this work follows a structured narrative that progressively builds upon the foundational concepts to explore advanced topics in consensus mechanism design and SSF realization. Chapter 2 explains the requisite background knowledge to understand distributed consensus and single slot finality. Chapter 3 explores the desired properties of Ethereum for SSF. Chapter 4 explores pragmatic mechanisms for achieving SSF in Ethereum via quantum signatures, core consensus-level changes, application-layer extensions, and sacrificing some level of economic security. This analysis explores fundamental tradeoffs between Ethereum's current properties and those desired for SSF. This thesis then systematically studies fast finality protocols through three families -- Propose-Vote-Merge, PBFT-inspired, and Total Order Broadcast/Graded Agreement. Their capabilities and limitations vis-à-vis Gasper are critically analyzed in Chapter 5. Additionally, Chapter 6 assesses the integrations and alternations necessary in D'Amato and Zanolini's recent work on a single vote TOB to cement SSF or similarly fast finality. The discourse culminates in the documentation of cumulative finality constructions and approaches in Chapter 7 to fortify the network's security. Conclusions and recommendations for future work are elucidated in Chapter 8. 

By assessing myriad approaches holistically, this thesis contributes vital groundwork, principles, and recommendations to guide the Ethereum protocol in prudently furthering SSF objectives. The suggested focus areas constitute an incremental research blueprint for decentralized consensus as blockchains scale.
\chapter{Background Knowledge}

In order to understand the current state of distributed consensus mechanisms, it is beneficial to have a historical background of prior works and concepts related to achieving distributed consensus.

\section{State Machine Replication}

A state machine is a computational model that encapsulates the dynamic behavior of a system~\citep{Nayak2019ConsensusSMR}. It maintains a system state and, when receiving inputs, transitions between states deterministically, producing an output. These transitions are reproducible, as the same protocol rules govern all transitions. This concept is pivotal to blockchains, whose ledger can be considered a state machine spread across multiple nodes.

State machine replication (SMR) is a foundational technique in distributed systems computing to ensure consistency, availability, and fault tolerance~\citep{implementing_schneider_1990}. The idea is to replicate the same state machine across multiple nodes or servers, traditionally called replicas, and ensure that every replica processes the same series of inputs in the same order. It is primarily concerned with handling non-malicious failures like crashes or network issues.

A critical challenge of SMR is to ensure that all replicas process inputs and state transitions in the same order. The deterministic nature of state machines is critical, as all nodes who receive the same input, given a common state, must transition to the same output to ensure consistency. A common approach is to use a consensus mechanism. Specifically, to protect against both crash failures and Byzantine failures, Byzantine Fault Tolerant consensus mechanisms are employed.

\section{Byzantine Fault Tolerance}

Byzantine Fault Tolerance (BFT) is a term used to describe consensus mechanisms that allow a distributed system to come to an agreement in the presence of malicious, or Byzantine, nodes. The term comes from the Byzantine Generals Problem, a thought experiment postulating how multiple divisions of an army, each with its own general, could communicate about agreeing on a time to attack a city~\citep{10.1145/357172.357176}. Communication can only occur via messengers; traitorous generals may send conflicting or inaccurate messages.

Analogously, distributed systems and their numerous nodes may experience a failure or be compromised by malicious actors. Ensuring a distributed system's correct and continuous function, even in the presence of these Byzantine nodes, is essential for security and reliability.

Two fundamental requirements for a BFT protocol are safety and liveness~\citep{practical_castro_1999}. Safety means that the honest nodes will not reach a wrong or conflicting consensus and satisfies linearizability, meaning the system's overall behavior is indistinguishable from a single, reliable, sequentially consistent system. Liveness is the ability to continue functioning and progressing the network state in the presence of Byzantine nodes instead of halting.

Both safety and liveness can only be guaranteed when there is an upper bound on the number of Byzantine nodes in the system. Historically, less than 1/3 of Byzantine nodes are tolerated in a system to achieve both properties simultaneously in the partially synchronous or asynchronous setting~\citep{fischer1986easy}.

Without this bound, messages from faulty or malicious nodes could outnumber honest nodes and control the network state.

\section{Practical Byzantine Fault Tolerance}

Practical Byzantine Fault Tolerance (PBFT) was published in 1999 and was the first state-machine replication protocol capable of surviving Byzantine faults, i.e., satisfying safety, in asynchronous networks~\citep{practical_castro_1999}. Before this work, the BFT problem was mainly theoretical due to high communication and computational overhead. An asynchronous network means there are no bounds on message delivery times in the network, and some messages may never be delivered. An assumption of synchrony is required for liveness, meaning the delay between message sending and receiving does not grow indefinitely. Otherwise, the protocol would provide safety and liveness in an asynchronous environment, which violates the availability-finality dilemma, an extension of the CAP theorem~\citep{lewispye2020resource}.

As long as less than 1/3 of nodes, or replicas as they are called in the paper, are faulty, the protocol functions as intended and provides both safety and liveness.

Nodes move through a series of states called views. In each view, one node is primary and the others are backups. A view change occurs when the primary node appears unresponsive. The significance of this feature is that it allows progress to be made continuously in the case of one node appearing faulty, with only a slight delay and no manual intervention needed.

Briefly, the system's basic structure involves a client requesting an operation to the primary node, that node multicasting the request to the backup nodes, the request being executed, and the response sent to the client. The client waits for \(f+1\) replies with the same result, where \(f\) is the number of potentially malicious or faulty nodes.

As discussed in~\cite{practical_castro_1999}, pre-prepare, prepare, and commit phases order requests across nodes. In the pre-prepare phase, a client request is assigned a number and is sent to the backups to accept. At this point, the backups multicast the prepare message; if they do not accept, which could happen if another request already has the same number, they do nothing.

After a node receives \(2f\) prepare messages from backups, the functional and honest replicas are guaranteed to agree on a total ordering for the requests within a view. A node then multicasts a commit message, and a variable representing if a given node commits to a message, the committed-local property, becomes true once a node gets \(2f+1\) commits. This phase and local commitments ensure that nodes agree on the numbering of requests, even if they occur in different views for each node. In blockchain-based SMR protocols, this prevents the concept of a double-spend, where one client can send the same transaction twice and get two separate results.

After committed-local is true, nodes execute the client's requested operation and send a result directly to the client.

Messages in this system are authenticated using cryptographic digital signatures to ensure that messages are arriving from their stated sender.

PBFT laid the groundwork for practical distributed systems that would one day become the basis for blockchain-based consensus mechanisms like Tendermint and HotStuff~\citep{10.1145/3404512.3404522}.

\section{Synchrony, Partial Synchrony, and Asynchrony}

A network's message delivery timing assumptions are critical when creating and analyzing consensus protocols. Different synchrony models lead to different safety/liveness guarantees and impossibility results.

A synchronous network assumes a known upper bound on message delivery time for all time. Since messages must be delivered within a specific time frame, this model is generally the easiest to reason about and can help determine which nodes are online and offline. In practice, a fully synchronous environment is a strict assumption and rarely holds in real-world networks where delays are inevitable and variable.

Asynchronous networks make no timing assumptions, meaning messages may be delayed or never delivered. As formalized in the FLP Impossibility, protocols can only achieve two out of three of fault tolerance, liveness, and safety properties in the asynchronous model~\citep{10.1145/3149.214121}.

~\citet{10.1145/42282.42283} introduced partial synchrony, which lies between these two extremes. It assumes an unknown Global Stabilization Time (GST). Before GST, the network is in an asynchronous-esque state where messages may be delayed arbitrarily. After GST, messages are delivered in a fixed time bound, as they are in synchronous environments. A GST of 0 means the network is always synchronous. Conversely, a GST of \(\infty\) means the network is always asynchronous. By considering this model, protocols must tolerate both periods of asynchrony and synchrony - a realistic approach to proving both safety and liveness and one commonly employed in consensus protocols.

\section{Nakamoto Consensus/Bitcoin}

Bitcoin, released in 2008, was the first implementation of a blockchain and the Byzantine Fault Tolerant consensus mechanism now colloquially known as Nakamoto Consensus~\citep{bitcoin_nakamoto_2008}.

This model varies from the direction of traditional BFT research and introduces the notion of dynamic participation. In previous mechanisms like PBFT, nodes were assumed to be continuously available and involved in consensus. On the contrary, the Nakamoto Consensus model allows nodes to join and leave the network an arbitrary amount at an arbitrary time without being considered adversarial to the network. This dynamic availability better mirrors real-world scenarios where network nodes are subject to unpredictable downtimes, maintenance, or other outages. Consequently, the robustness of the Nakamoto Consensus in accommodating such real-world irregularities makes Bitcoin's blockchain a more resilient and adaptable system.

In blockchains, transactions are bundled in increments called "blocks," which link to the previous bundle through a hash, creating a chain of blocks. Bitcoin uses a process called proof of work (PoW) for consensus. Miners are nodes in the network that compete to solve a cryptographic puzzle by brute force, generating a hash with a certain number of zeroes for every block. Each block has a header that contains information about the previous block hash, the Merkle root of transactions, the block timestamp, and the nonce. Miners increment the nonce to create different hashes. The protocol adjusts the puzzle's difficulty after every 2016 blocks, or approximately every two weeks, to maintain a stable network with new blocks approximately every 10 minutes, even as the total computation power of miners changes~\citep{10.1007/978-3-030-03577-8_41}. The difficulty is adjusted by varying the leading zeroes required in the SHA-256 hash of the block's header.

When a miner finds the correct nonce and thus a valid hash for the block, it broadcasts this block to all other miners, who only accept it if the transactions adhere to the protocol rules, such as not transferring more Bitcoin than one owns. If accepted, the miners then begin attempting to create the next block in the chain and use the hash of this new block as the previous hash. By adding a new block, a miner receives a block reward in the form of new Bitcoins that come from built-in protocol inflation that decreases over time and transaction fees from the transactions included in the block. This economic incentive and the opportunity cost of the mining hardware and its operation encourage miners to act honestly.

If two different miners find a solution to the same block simultaneously, the network will adhere to whichever block becomes part of the longest chain. In practice, this looks like one-half of the network working to create a block using the previous block hash from miner A and the other half using the previous block hash from miner B. One of these groups will eventually find a solution and broadcast it to the whole network, at which point all miners switch to this longest chain.

To bolster the network's resistance against potential vulnerabilities and transaction reversals because of a switch in the longest chain, Bitcoin clients and recipients have adopted the $\kappa$-deep confirmation rule. This rule necessitates waiting for a transaction to be embedded $\kappa$ blocks deep in the blockchain before confirming it. Six confirmations, meaning six blocks deep, are typically awaited before recognizing a transaction as irreversible. From a safety perspective, the deeper the transaction is in the chain, the more computationally challenging and economically infeasible it becomes for an adversary to alter or reverse it. However, even with the 6-deep confirmation rule, an adversary controlling 30\% of the mining power still has an attack success probability of 17.74\%, meaning there is a 17.74\% chance the attacker could double-spend coins~\citep{bitcoin_nakamoto_2008}.

With only probabilistic finality, Nakamoto Consensus carries a potential vulnerability: the 51\% attack~\citep{Pan2018/05}. Suppose a miner, or a cabal of miners, controls over 51\% of the network's computational power. In that case, they will have the highest probability of finding new blocks, leading to a monopoly on the blockchain's growth and transaction proliferation, bringing the possibility of a double-spend attack. To start the attack, they will secretly mine blocks and not broadcast them to the rest of the network. While doing this, they will send a transaction using their Bitcoins on the current longest chain. Once the transaction has received the standard six confirmations, the malicious miner reveals their secret fork, where the transaction has not occurred. Since this malicious actor holds the majority of the computing power, it will eventually catch up to and surpass the length of the longest chain, leading all other miners to adhere to it going forward. Now, in the canonical view of the chain, the miner's transaction never happened, and they still have their coins. However, the transaction recipient may have already provided a good or service in exchange, allowing the miner to double-spend the same coins more than once.

Nakamoto Consensus represented a groundbreaking shift in distributed systems and consensus algorithms. Future works on blockchain-based BFT consensus mechanisms, many of which will be covered in this work, have built upon the concept of dynamic participation and achieved stronger finality guarantees.

\section{Sleepy Model}

One of the most transformative aspects of Nakamoto Consensus was its inherent support for dynamic participation, e.g., allowing nodes to freely join and leave the network without penalties or predefined commitments. This concept was refined and formalized in the Sleepy Model~\citep{sleepy_pass_2017}. It defines the construction of protocols where the central assumption is that a majority of \textit{online} nodes are honest, instead of a majority of all nodes.

Nodes are allowed to unpredictably "sleep," or go offline, and subsequently "wake up," coming back online. Unlike previous consensus models, where offline nodes were considered faulty or malicious, the sleepy model accepts these dynamics as expected behavior in large decentralized networks. This model provides a much more realistic view of distributed systems.

The sleepy model also removes the need for proof of work. However, its analysis in~\citet{sleepy_pass_2017} takes place in the permissioned setting instead of the permissionless setting, meaning there is a fixed set of participants instead of anyone being able to join. Future works adapted the sleepy model to the permissionless setting using proof of stake~\citep{damato2023recent}.In short, the adaptation uses a random oracle, typically a pseudorandom function like a hash function, and uses a node's cryptographic identification information and the block-time as inputs. The nodes that produce a hash lower than a set baseline become leaders for that block proposal.

\section{Quorum Certificates}

Quorum Certificates (QCs) prove that a certain fraction of validators in a distributed system have sent the same message about a value or state, typically attesting or agreeing to its validity~\citep{sync_abraham_2020}. Some protocols require a QC before a value is committed or finalized. To create a QC, validators must first gather signatures. Once they have collected enough signatures, these can be combined and compiled into a QC.

Compared to elementary message sending, QCs provide greater efficiency by removing the need for nodes to communicate messages across the entire network. However, they traditionally require a static participation group, as the participation threshold is based on the number of active validators, and a dynamic participation model may result in each validator having a different view of who is awake.

\section{Availability-Finality Dilemma and Ebb-and-Flow Protocols}

Recall that a dynamically available protocol can preserve safety and liveness as long as the majority of online validators are honest. As is implicit in the construction of Nakamoto Consensus, these protocols are not tolerant to network partitions. For example, suppose the Bitcoin network splits due to a network or latency issue. In that case, nodes on both sides will continue per usual under the assumption that the other nodes are simply offline, and there will be two chains, each claiming to be the longest until they can communicate with each other.

There is no risk of network partition in a permissioned environment, where the number of nodes is fixed and dynamic participation is not tolerated. Instead, if over a certain number of nodes appear offline, the protocol will halt as it cannot reach a consensus without a quorum of online users. This problem relates to the distributed computing CAP theorem, which states that a distributed data store can provide, at most, two of the three guarantees of consistency, availability, and partition tolerance~\citep{10.1145/343477.343502}. This theorem was brought to blockchain consensus and analogized consistency to security and liveness to availability~\citep{lewispye2020resource}. This availability-finality dilemma, as it was called, means that no one ledger can provide both dynamic participation and deterministic, irreversible finality - a choice must be made between dynamic participation and liveness or safety and network partition resilience.

Consequently,~\citet{ebbandflow_neu_2021} considered the case where individual clients may want different safety and liveness tradeoffs. They proposed a formalization of the design seen in Ethereum's current Gasper protocol, where there are two ledgers: one for liveness and dynamic availability, and another, a prefix of the first, for safety and finality under network partitions. These ebb-and-flow protocols allow for a consistently live protocol to always be making progress. At the same time, another ledger with stronger safety guarantees can give deterministic finality instead of Nakamoto-style probabilistic finality.

This idea was also expanded into the snap-and-chat protocols, which achieve ebb-and-flow properties while having the optimal tolerance for both safety and liveness~\citep{neu2020snapandchat}. Less than 50\% of validators can be malicious to guarantee liveness and less than 33\% for safety. This construction is achieved by requiring nodes to execute both protocols simultaneously, with the safe ledger taking occasional snapshots from the available ledger for finalization.

Clients can choose the safety-liveness tradeoff they want to make according to their use and risk tolerance.

\section{Gasper}

Ethereum's current consensus protocol, Gasper, is in this ebb-and-flow family of protocols~\citep{combining_buterin_2020}. Its dynamically available chain follows the output rule of the LMD-GHOST fork choice rule, and the Casper Friendly Finality Gadget (FFG) manages the finalized chain.

Unlike Nakamoto consensus, Ethereum utilizes proof of stake, where a node or validator's network influence is proportional to their bonded stake in the system. To become a validator eligible for proposing and attesting blocks, 32 ETH, the native cryptocurrency of the Ethereum blockchain, is staked or put up as collateral~\citep{grandjean2023ethereum}. If a validator acts malicious and violates one of two slashing conditions by attempting to double-vote or change the organization of blocks, their stake is destroyed. On the other hand, honest participation and block proposal rewards a validator with native inflation and transaction fees. This system creates economic incentives to validate honestly, similar to Nakamoto consensus, without requiring powerful computing hardware and operating costs.

Gasper proceeds in slots and epochs. During each slot, a block is proposed by a randomly picked validator. There are 32 slots per epoch. The validators are partitioned into committees in each slot, with one committee per slot and no validators in more than one committee per epoch. During each slot, validators have one of two roles:

\begin{enumerate}
    \item Propose a new block to be appended to the LMD-GHOST chain
    \item Vote for the canonical chain that aligns with their view of the LMD-GHOST fork choice, and vote for FFG Casper to justify and finalize the latest epoch boundary pairs.
\end{enumerate}

These messages and proposals get broadcasted over a peer-to-peer network.

\subsection{LMD-GHOST}

Latest Message Driven Greediest Heaviest Observed SubTree, or LMD-GHOST, is a greedy algorithm fork-choice rule that allows nodes or validators to attest to a block and make decisions about the current canonical available chain~\citep{combining_buterin_2020}. This fork choice method allows validators to choose chains based on the votes, or weight, of its subtree. At any given time, validators will conclude that the subtree of blocks with the heaviest weight is the correct canonical tree. The "LMD" in LMD-GHOST signifies that only the most recent set of attestations from each validator is regarded in the fork choice, ensuring that only the most updated state is considered by validators. These properties allow it to preserve dynamic participation and liveness.

An LMD-GHOST chain is only capable of probabilistic finality at best, meaning it could be the victim of several different attack vectors that result in a chain reorganization and a double spend. For example, a short-range attack could occur, where a malicious actor uses previous voting power to create an alternate chain history that appears to be the preferred option with the most weight to new nodes~\citep{schwarzschilling2021attacks}. This attack vector is mitigated in longer periods by having a finality gadget like Casper and by having new nodes join the network from a weak subjectivity checkpoint. Still, the ability to use this attack on yet-to-be-finalized blocks remains.

Another is the balancing attack, wherein an adversarial block proposer could produce two conflicting blocks revealed to equal-sized subsets in the voting committee~\citep{neu2022attacks}. Then, in the next slot, it could release the withheld votes from the previous slot to split the following committee's voters into two subsets, each with a different view of which chain is leading and voting for that one. This process, if repeated indefinitely, causes liveness to crumble. Ethereum's version of LMD-GHOST attempted to patch this vulnerability by introducing weights on proposals, but~\citet{10.1145/3560829.3563560} demonstrated that the balancing attack is still possible.

Ethereum's implementation of LMD-GHOST uses subsampling, meaning only a portion of the entire validator set votes on each block as part of their committee during an epoch. A significant vulnerability under this model is the ex-ante reorg attack, which becomes feasible under certain conditions~\citep{neuder2021lowcost}. This attack scenario begins when an adversary, controlling a $\beta$ fraction of validators and including the proposer for slot $t+1$, secretly creates a block $B'$ atop the existing block for slot $t$, $B$. If this adversary controls roughly $\beta$ of the validators chosen to vote at slot $t+1$, these validators can covertly vote for $B'$. Meanwhile, honest voters, unaware of $B'$'s existence, cast their votes for $B$. In the subsequent slot, $t+2$, an honest proposer, basing their actions on $B$, releases another block, $B''$. Assuming they are honest, the validators of this slot will naturally vote for $B''$, giving it a weight of $(1-\beta)W_c$.

The attack unfolds further when the adversary reveals $B'$ alongside the votes accumulated during slots $t+1$ and $t+2$. With this revelation, $B'$ attains an approximate weight of $2\beta W_c$. This weight becomes crucial as, if $\beta$ exceeds $1/3$, $B'$ usurps $B''$'s position as the canonical block. The attack's potency is amplified if an adversary controls $k$ consecutive slots, a scenario with a probability of $\beta^k$. In such a case, the adversary can withhold votes from these slots, only to reveal them after an honest slot intervenes. This strategy can accumulate a weight of $(k+1)\beta W_c$ for an adversarial block. A reorganization of the honestly proposed block occurs if $\beta$ is greater than $1/(k+2)$.

The role of subsampling in this attack cannot be overstated. It is the mechanism that allows the adversary to accumulate votes from controlled validators across k committees. In its absence, votes from the last slot would merely overwrite the previous ones, courtesy of the latest message rule. Moreover, without subsampling, an adversary's capacity to exceed the weight of all honest validators voting collectively in a single slot is significantly hindered, given the majority of honest validators in the overall set.

\subsection{Casper FFG}

Casper the Friendly Finality Gadget (Casper FFG) is an overlay designed to add finality guarantees to any blockchain~\citep{casper_buterin_2017}. Inspired by PBFT, it has justification and finalization steps analogous to PBFT's prepare and commit phases.

Since it is a finality overlay, there is no notion of block proposers, and it justifies and finalizes the proposal mechanism's block tree.

Every block has a height, representing its distance from the genesis block. Checkpoint blocks are those whose height is a multiple of a constant value. Casper's role is to justify and finalize these checkpoint blocks and create a "checkpoint tree." These checkpoints are aligned with the beginning and ends of epochs and denoted as epoch boundary blocks. More specifically, Casper considers epoch boundary pairs, which are pairs of a block B and epoch j, denoted (B,j). This consideration clarifies situations where B is an epoch boundary block in some chains but not others due to forks.

Let $LJ(\alpha)$ represent the latest justified pair of attestation $\alpha$, the highest attestation epoch justified pair in the view of the latest epoch boundary block. Similarly, let $LE(\alpha)$ be the last epoch boundary pair of $\alpha$.

In each slot, validators publish an attestation with $LJ(\alpha) \rightarrow LE(\alpha)$.

In a validator's view, epoch boundary pair (B,j) is justified by (A,j') if at least 2/3 of the total validator stake has voted for an attestation of (A,j') $\rightarrow$ (B,j). When a 2/3 majority votes for an attestation, it is known as a supermajority link. Epoch boundary pair (A,j') is finalized if it is justified, (A,j') $\rightarrow$ (B,j) is a supermajority link, and B is the epoch boundary pair immediately after A.

Casper FFG validators can have their stake slashed if they violate one of two conditions. The first is if they vote for two checkpoints at the same block height. The second is if a validator surround votes, meaning they have votes A $\rightarrow$ B and C $\rightarrow$ D, but C.timestamp $<$ A.timestamp $<$ B.timestamp $<$ D.timestamp.

During every epoch, validators run the fork choice rule and make an attestation. Once a block is finalized, it is secure regardless of network latency or temporary asynchrony. To reorganize finalized blocks, over 1/3 of validators would have to collude in a way that would destroy their stake, equivalent to 1/3 of the network's stake. This gives Casper, therefore Gasper, the property of accountable safety and economic finality. Additionally, as long as the underlying consensus protocol creates blocks, new checkpoints can be justified and finalized, giving the protocol plausible liveness.

\chapter{Desired Properties for Single Slot Finality on Ethereum}

In the current Gasper architecture, Casper FFG does not confirm blocks until between 64 and 95 blocks, approximately 15 minutes after they appear in the LMD-GHOST chain~\citep{damato2023simple}. Single slot finality (SSF) is the concept of blocks being proposed and finalized in the same slot. There are several reasons for this to be a desired trait, namely:

\begin{itemize}
    \item Faster confirmations: The user experience is greatly improved by having transactions almost immediately confirmed instead of requiring a 15-minute wait.
    \item Reorg resilience: With immediate finality, there is no risk of a short-term block reorganization, which could happen if validators are attempting to censor a specific block or extracting value from transaction ordering (MEV)~\citep{KonstantopoulosButerin2021}.
    \item Stronger security guarantees: After a block is proposed and finalized, removing that block from the network would require 1/3 of the staked ETH to be slashed. Additionally, since achieving SSF would likely require replacing LMD-GHOST, it would mitigate LMD-GHOST's attack vectors like ex-ante reorgs and balancing attacks.
\end{itemize}

Though consensus mechanisms like PBFT have previously been modified to achieve instantaneous finality on a blockchain (e.g., Tendermint), specific desired characteristics for an LMD-GHOST replacement make these alternates insufficient~\citep{latest_buchman_2018}.

\section{Drop-in Replacement for LMD-GHOST}

Given that Ethereum is already a live, highly-adopted network, making the transition between consensus mechanisms should be as seamless as possible. The new SSF-compatible consensus mechanism or fork choice rule should be compatible with the existing Ethereum architecture. Namely, it should be similar to LMD-GHOST, compatible with Casper FFG, and simple enough not to require burdensome computational overhead. Additionally, its round latency should not be significantly higher than the current $\sim12$ seconds, as otherwise, some of the benefits of SSF are lost.

\section{Subsampling}

Similarly, the protocol would ideally be compatible with validator subsampling. Currently, each epoch has distinct sets of 32 committees, one per slot, further divided into 64 subcommittees ~\citep{ButerinSingleSlotFinality}. Assuming there are $2^{19}$ validators, each subcommittee has about 256 validators. Each committee has sixteen committee aggregators, which are chosen by the network and are responsible for combining as many signatures from the committee as possible into a single signature and publishing it to the main peer-to-peer subnet. The block proposer chooses the aggregated signature with the most participants by stake-weighted balance.

Since Casper FFG finalization requires attestations from 2/3 of the staked validators, achieving this level of voter participation will likely require efficient aggregation and no subsampling. Research is currently being conducted to find a better aggregation scheme, which may include adding another layer of subcommittee aggregation or zero knowledge aggregation techniques~\citep{Kadianakis2023SignatureMerging}. Any more complicated scheme would require longer slot times to account for extra latency and involve more complexity. However, it may be worth the benefits of simple, economically secure single slot finality and the entire set of validators voting in each slot. Section 4.4 will discuss the possibility of accepting fewer signatures and its consequences.

\section{Asynchrony Resilience/Dynamic Availability}

For our protocol to handle the uncertainties of real-world scenarios, it must have some level of temporary asynchrony resilience. Otherwise, a temporary stall in liveness could compromise the chain and all previous finalizations. As we know from the CAP Theorem and the availability-finality dilemma, a protocol cannot be both safe under dynamic participation and live under network partitions or temporary asynchrony~\citep{lewispye2020resource}. At the same time, we would like dynamic availability - for both safety and liveness to be guaranteed while dynamic participation is possible. Ebb-and-flow protocols, like Gasper with its available and safe chain, are used to counter this problem~\citep{ebbandflow_neu_2021}. Given that we are only interested in replacing LMD-GHOST, not Casper FFG, the safe chain requirement will be inherently satisfied, and we are concerned with providing the mechanism by which transactions can be finalized in the slot in which they were proposed.

\chapter{Methods to Achieve Single Slot Finality}
Various methods exist to introduce single-slot finality into the Ethereum blockchain, ranging from the protocol to the application level. This section will investigate and evaluate the available options based on their feasibility and complexity. 

\section{One-Shot Quantum Signatures}
Recent advances in quantum cryptography have led to the introduction of a new type of digital signature called a one-shot signature~\citep{cryptoeprint:2020/107}. These signatures use quantum physics to create private keys that can only sign a single message. The keys exist as quantum states, meaning they cannot be copied or cloned due to the quantum no-cloning theorem~\citep{Park1970-PARTCO-16}~\citep{Ben_David_2023}. 

In quantum physics, particles do not have singular defined properties, like position, prior to measurement. Instead, they exist in a superposition of multiple probable states. The act of measuring forces the quantum state to "collapse" into a single definite value. Analogously, the private keys existing as quantum states exist in a superposition of many different signature keys. There is no singular key to copy, hence the no-cloning theorem. 

Producing a signature using the key forces the superposition to collapse into a single signature key, which is useless for anything else. The key cannot be reused because the superposition is destroyed.

One-shot signatures may allow for provable single-slot finality in Ethereum. The idea is to have validators create chains of one-shot signatures~\citep{progcrypto_oneshot_signatures}. A validator will generate the quantum public-private keypair, sign a message, and include in the message the public key that will be used in the following signature. These signature chains could also enforce specific rules, such as requiring an epoch counter to be strictly increasing. Equivocation could be prevented by including source and target epoch counters and requiring that the source never decreases and the target always increases. 

Once a block achieves a majority (51\%) of attestations from validator one-shot signature chains, fork-free finality is cryptographically guaranteed. No future block can accumulate enough signatures to finalize an alternative fork. This method allows for incredibly strong cryptographic single-slot finality secured by the principles of quantum physics.

Unfortunately, one-shot signatures are still purely theoretical and likely decades away from an implementation that could work in this setting. Validators would require access to a quantum computer - a massive obstacle to overcome.

\section{Core Consensus Change}
The most likely path to long-term single slot finality is through a change to the Ethereum fork-choice rule, LMD-GHOST. With a different fork-choice rule that supports fast confirmations, ideally within the same block, single-slot finality could be achieved.

The main downside to this approach is that it will likely take a few years for a replacement to be decided upon, rigorously tested and developed on Ethereum's numerous clients, and introduced via hard fork after gaining user agreement through multiple rounds of voting. Though this process ensures that the ultimate implementation is safe and agreed upon by the community of validators, it also means that Ethereum will be using an out-of-date fork choice rule that is difficult to perform a proper security analysis on, creating an opportunity for other blockchains to capture more significant market share.

\section{EigenLayer Actively Validated Service}
EigenLayer is an Ethereum restaking framework that enables validators to pool security across different blockchain modules, protocols, and applications~\citep{eigenlayer2023whitepaper}. Validators that opt to re-stake ETH into EigenLayer can provide validation services to Actively Validated Services (AVS) and earn additional returns beyond consensus rewards. In exchange, restakers are subject to slashing conditions imposed by the AVS modules they opt to serve. This pooled security model allows protocols and services to leverage Ethereum's crypto-economic guarantees, robust security, and decentralization. 

One possible application of EigenLayer is providing single slot finality for transactions or blocks. Under a basic SSF model, a set of validators re-staked on an EigenLayer AVS module could attest that they will never vote for a chain that reverses or excludes a block after it has been proposed. Validators restaking to this AVS would finalize the block in a single slot rather than waiting for probabilistic finality. However, several challenges exist with this approach:

First, a majority of Ethereum validators would need to re-stake into this EigenLayer SSF module for the guarantees to be meaningful. Otherwise, in the case of a reorg, the minority SSF validators may be forced to violate their attestations or face inactivity penalties from the Ethereum base layer for failing to build on the canonical chain. Reaching this threshold for a new opt-in module presents difficulties. 

Second, bolting SSF onto Ethereum in a minority way risks creating coordination failures, chain splits, or breaking consensus incentives if the minority SSF validators clash with the majority. Careful crypto-economic protections are needed to align incentives.

Third, providing SSF at the transaction level via tips or fees faces challenges around fairly compensating SSF restakers and block proposers without enabling MEV manipulation or requiring non-EigenLayer stakers to opt-in to some extra rule or client upgrade. Preventing oligopolistic behavior is also tricky in a minority SSF system.

While EigenLayer's pooled security model offers intriguing possibilities, implementing strong SSF guarantees for Ethereum without majority adoption or base layer modifications remains challenging. A better pathway may be using EigenLayer to facilitate fast finality sidechains, which is currently being developed~\citep{near_eigenlabs_2023}. If EigenLayer reaches a supermajority of validators, an SSF AVS could be enabled, and SSF could be implemented much more quickly and without changing the core protocol. This system could provide valuable data while limiting disruption risks to Ethereum's core consensus. It could be replaced with a proper SSF solution at the protocol level later.

\section{Sacrifice Full Sampling - A Philosophical Modification to SSF}
~\citet{buterin2023sticking} discussed the possibility of sacrificing the current finalization process, which requires 2/3 of the validator set to cast FFG votes, in favor of a more efficient but less secure alternative. Reducing the requirement to 8192 signatures mitigates some of the signature aggregation issues previously discussed. It ensures that Ethereum can future-proof its operation regardless of future validator set size or the need for a quantum-resistant signature aggregation mechanism.

Buterin proposed three approaches that sacrifice Ethereum's current high cost for equivocation and malicious behavior in favor of efficiency. The most likely feasible approach is to create accountable committees, with 4096 validators per slot with various amounts of ETH staked.

At a certain variable level of stake M, a validator would be guaranteed inclusion in the validator set, and all other validators would have an N/M chance to validate for a given slot. One attractive characteristic of this model is that rewards and consensus weight do not have to be proportional, allowing finality breakage to require more ETH to be malicious than individual actors.

The elegance of this approach comes from its ability to provide a variable amount of ETH staked per validator while still maintaining stringent slashing rules and appropriately rewarding and selecting validators. If implemented, it would make the primary obstacle to the implementation of SSF a consensus-related matter.

Though this approach would be beneficial in an ideal scenario, it remains to be seen if there would be enough incentive for validators with significant stake to centralize it into one validator. They may continue running several validator clients with smaller stakes to mitigate potential issues like downtime leading to inactivity leaks caused by more than one-third of the network being offline for an extended period - a formula currently based on a quadratic model~\citep{edgington2023inactivity}. The issue then becomes how to incentivize validators with lots of ETH to stake from one validator while still incentivizing general validator decentralization.

One possible way to solve this issue would be with non-linear penalty scaling for inactivity leaks. The main benefit of running one validator as opposed to several, besides the general maintenance and logistical simplification, is reducing the operational cost of running several nodes. Though simplification and validator cost savings could act as an incentive to run one validator instead of several for the time being, this will be mitigated as Ethereum further advances toward its goal of SNARKifying the EVM and making it possible to run validators on mobile phones~\citep{retford_drake_2023_ethvalidator}. By making the inactivity leak penalty smaller for larger nodes or monetarily equivalent to other sized nodes, the penalty of running one large node would be mitigated and, in fact, incentivized due to the lower operational costs. In practice, this could be done by weighing the inactivity score by the validator's stake or changing the penalty formula for the larger validators to have their inactivity leak follow the distribution of smaller validators. By integrating this mechanism, a single large entity could run one validator with all of its stake and be subject to the same inactivity leak as if one of many minimal-stake validators became inactive.

It may so happen that validators are comfortable with staking more ETH and taking on the risk of inactivity leaks for higher guarantees of inclusion in the validation process and reducing the capital cost of running multiple validators. The formal analysis and economics of modifying the inactivity leak for larger stakers are left to future work.

Another approach would be to add an incentive for being a large staker, possibly around more frequent block proposals. However, this may cause the unintended effect of pricing smaller stakers out of the market and creating large staker cabals that engage in censorship.

Ultimately, scaling down signatures causes a decrease in economic finality on Ethereum, sacrificing some of the system's security for an increase in operational efficiency. The community must decide upon a change of this philosophical magnitude, and it may be a contentious issue over the coming months and years.
\chapter{Protocol Overviews}

This section will explore several blockchain consensus mechanisms with various constructions. Some are considered potential replacements for Ethereum's LMD-GHOST implementation. In contrast, others are actively used in different blockchains and boast a form of single slot or instantaneous finality. These constructions were chosen for analysis to assist in understanding the current state of distributed mechanisms and what techniques could be utilized in an LMD-GHOST replacement. The overviews provided here focus on conveying these protocols' conceptual frameworks and operational characteristics rather than restating their formal proofs and algorithms. 

Each protocol's summary is an original interpretation intended for comparative analysis, and extensive citations to their respective original sources are included. Unless stated otherwise, these sources contain the full technical details and formal proofs of safety and liveness—the two main requirements for a secure and usable consensus mechanism. In line with academic integrity, this document upholds rigorous standards of citation and attribution to ensure that the ideas and contributions of original works are appropriately acknowledged. The critical evaluation and synthesis of these protocols within the context of SSF for Ethereum are a unique contribution of this thesis, aimed at contributing to the ongoing development and refinement of distributed consensus mechanisms.
\section{Organization of Consensus Mechanism Families}

The consensus mechanisms explored in this thesis are categorized into families based on lineage and structural similarities. This organizational approach facilitates a systematic and coherent study of each family's unique attributes and the individual mechanisms within them. The following families are central to our discussion:

\begin{itemize}
    \item \textbf{Propose-Vote-Merge Family}: This family, first characterized in~\citet{damato2023recent} encompasses fork choice protocols that primarily follow a propose-vote-merge pattern. Protocols in this category are characterized by their stages of a leader proposing blocks, active validators voting, and finally, validators using the view-merge synchronization technique to align with the proposer's chain view for the next slot. Since they are fork choice protocols, they are easily portable into the ebb-and-flow architecture of Ethereum and Casper FFG. Examples include the Goldfish and RLMD-GHOST protocols, each discussed in its dedicated subsection.
    
    \item \textbf{PBFT-inspired Family}: Protocols in this family derive from the Practical Byzantine Fault Tolerance (PBFT) algorithm, known for its early application and implementation in distributed computing~\citep{practical_castro_1999}. They are designed to offer immediate finality and typically involve a sequence of pre-prepare, prepare, and commit stages to reach consensus. This family includes adaptations like Tendermint and HotStuff, which have modified the original PBFT to suit blockchain environments~\citep{latest_buchman_2018}~\citep{hotstuff_yin_2019}. Like PBFT, they typically operate with a fixed set of validators and no dynamic participation, but some have modifications to enable some level of participant joining and leaving. With little to no support for dynamic participation, if there are not enough validators to reach a quorum, the mechanism may halt, preserving safety over liveness. Partly as a consequence of this tradeoff, these protocols are generally able to provide instant finality.
    
    \item \textbf{Total-Order-Broadcast/Graded Agreement Family}: Total-Order-Broadcast (TOB) and Graded Agreement (GA) protocols focus on achieving a total order across distributed processes in the presence of Byzantine faults. Graded agreements are a weak form of consensus used as a building block for complete consensus protocols~\citep{10.1145/62212.62225}. They are key for ensuring that participants achieve consensus on the order of messages, even with Byzantine faults, by enabling a structured decision-making process where validators assign grades to proposals based on their support level from other validators. Protocols like Momose-Ren and D'Amato-Zanolini, which aim to refine and build upon the fundamental concepts of TOB and GA, fall into this family~\citep{constant_momose_2022}~\citep{damato2023streamlining}.
    
\end{itemize}

Following this overview that outlines its general principles and significance in the broader context of distributed consensus, subsections dedicated to individual protocols will delve into their operational details, highlighting how they address the challenges of safety, liveness, and finality. The analysis will also consider their applicability to Ethereum's vision of SSF, scrutinizing the potential each has to enhance or replace the current LMD-GHOST protocol. Unless otherwise stated, \(\Delta\) is an upper bound on network delay used to explain the latency of some protocols.

\section{Propose-Vote-Merge Family}
\subsection{Goldfish}
\citet{damato2023goldfish} authored Goldfish as a potential replacement for Ethereum's LMD-GHOST. Its innovative use of message buffering is its pivotal contribution to single slot finality consensus mechanisms. The idea of precisely timing the inclusion of network votes into the external pool through message buffering ensures that all honest validators view a slot proposer's perspective as both complete and valid, compelling them to vote in favor of the proposer's block. This timing is achieved by allowing the proposer to be the last to update its message buffer and guiding other validators to synchronize their local views with the proposer's. This concept, termed as view-merge, is integral to achieving reorg resilience and preventing exploits like balancing attacks.

Goldfish operates in a time-divided manner, with slots extended to \(3\Delta\) to ensure synchronous operation. Supporting subsampling, each slot is divided into a committee of active validators, and does not require the entire set of validators to vote every slot.

\subsubsection{Protocol Phases}
Goldfish consensus is segmented into three phases:

\begin{enumerate}
    \item \textbf{Propose Phase}:
    \begin{itemize}
        \item At the outset of a slot, the proposer chosen by the verifiable random function integrates its buffer with its block-vote tree (bvtree).
        \item The proposer then runs its fork choice rule and proposes a new block at the tip of the chain.
    \end{itemize}
    \item \textbf{Vote Phase}:
    \begin{itemize}
        \item Validators synchronize their bvtree with the proposer's view, ensuring uniformity in view across the network.
        \item Post-synchronization, validators eligible to vote in that slot cast their votes based on the fork-choice function and last slot's votes. 
    \end{itemize}
    \item \textbf{Confirm Phase}:
    \begin{itemize}
        \item Validators consolidate their buffers into their bvtree.
        \item Validators identify the tip of the chain using votes from the current slot and output as the confirmed ledger the transactions from slots \(\kappa\)-deep in time.
    \end{itemize}
\end{enumerate}

The view-merge ensures that all honest validators have a unified view of the state of the network, thereby achieving reorg resilience, meaning that blocks submitted by honest validators are eventually guaranteed inclusion in the blockchain.

\subsubsection{Features and Observations}
\paragraph{Vote Expiry}

Unlike LMD-GHOST, where votes have no expiration and many different attacks are possible, Goldfish employs a strict 1-slot vote expiry period, meaning that only the last round's votes are considered in the fork choice function for finding the current canonical chain. This rule leads to the reorg resilience property and the impossibility of ex-ante reorgs under synchrony. If all honest voters in one slot voted for a particular subtree, the next slot will also vote on the same subtree.

\paragraph{Fast Confirmation}

Goldfish also supports the inclusion of an additional fourth phase, fast-confirm, that allows for constant expected confirmation latency regardless of security level under high participation and an honest supermajority.

Going between the vote and confirm phase, fast-confirm involves the validator merging its buffer into its bvtree and marking a block as confirmed if the number of votes for block \(B\) in the current slot is greater than or equal to \(\frac{3n}{4} + \frac{\epsilon n}{2}\) for some \(\epsilon > 0\). The confirm phase will output the higher of the fast-confirm chain and the \(\kappa\)-deep prefix.

\paragraph{Compatibility with Finality Gadget}

Goldfish is meant to replace LMD-GHOST in Ethereum and is the dynamically available chain in the ebb-and-flow construction, where Casper provides the accountably safe chain. Like the current construction, the Casper chain would be a prefix of the Goldfish chain and, as input, would take confirmed blocks and subtrees.

\paragraph{Temporary Asynchrony}

Due to Goldfish's strict 1-slot vote expiry, any temporary asynchrony could be catastrophic to the protocol's safety \citep{damato2023goldfish}. For example, after the 1-slot expiry, any votes previously cast for block \(B\) will no longer be valid, and the reorg resilience property is completely lost. In practice, this makes Goldfish unacceptable as a replacement for LMD-GHOST.

\subsubsection{Remarks}

Though Goldfish marks a significant step towards an SSF-compatible fork-choice mechanism for Ethereum, its vote expiry proves too strict to operate in a real-world environment. The risk of temporary asynchrony is far too great to justify the 1-slot expiry.


\subsection{RLMD-GHOST}
RLMD-GHOST is a consensus protocol designed to support dynamic participation while allowing for a bounded tolerance of temporary asynchrony \citep{damato2023recent}. Building upon the techniques of the Goldfish protocol, it modifies the strict 1-slot vote expiry to provide greater safety and flexibility in environments with temporary asynchrony while still retaining support for view-merge and some level of reorg resilience.

Operating within the partially synchronous model, RLMD-GHOST divides time into slots of length \(3\Delta\). The protocol does not utilize committees to prevent possible ex-ante reorgs that could occur with votes accumulating from multiple committees over time, meaning that every awake validator votes in every slot.

\subsubsection{Protocol Phases}

RLMD-GHOST progresses through cycles of Propose, Vote, and Merge phases:

\begin{enumerate}
    \item \textbf{Propose Phase}: At the start of a slot, the proposer merges its view with its buffer and broadcasts the proposed block.
    \item \textbf{Vote Phase}: Validators synchronize with the proposer's view and broadcast their votes for the local chain head.
    \item \textbf{Merge Phase}: Validators merge their views and buffers, solidifying the chain state.
\end{enumerate}

\subsubsection{Features and Observations}
\paragraph{Vote Expiry}
A pivotal innovation in RLMD-GHOST is the introduction of the vote expiry period parameter \(\eta\). Contrasting with Goldfish's 1-slot vote expiry, RLMD-GHOST considers only votes from the most recent \(\eta\) slots as valid within the fork choice function. This adaptation ensures a balance between fast finality and resilience to network asynchronies.

\paragraph{Generalized Sleepy Model}

The generalized sleepy model, in which RLMD-GHOST operates, extends the standard sleepy model by introducing a parameter \(\tau\) that signifies the "sleepiness" coefficient~\citep{damato2023recent}. A protocol is \(\tau\)-dynamically available if it upholds safety and liveness under the \(\tau\)-sleepy model, where the honest validator set from slots \(t-\tau\) to \(t-1\) outweighs adversarial validators in slot \(t\).

RLMD-GHOST maintains \(\eta\)-dynamic availability for any \(\eta \leq \tau\), underpinning its asynchrony resilience. This capacity ensures safety continuity when the honest validators from \(\eta\) slots prior overpower current adversaries and inactive validators.

\paragraph{Fast Confirmation Phase}

In addition to the standard phases, RLMD-GHOST also introduces a fast confirmation phase for rapid block finalization under optimal conditions. This phase occurs between the Vote and Merge phases and aims to achieve confirmation latencies akin to those of traditional BFT protocols. A validator can note a block as confirmed as soon as it receives votes from 2/3 of validators instead of waiting for the latency timeout. This does require the optimistic assumption that 2/3 honest validators are awake. When this is not the case, RLMD-GHOST reverts to the slower $\kappa$-deep confirmation rule that is live under dynamic participation.



\paragraph{Single Slot Finality Protocol}
\citet{damato2023simple} builds upon RLMD-GHOST to introduce modifications aimed at making RLMD-GHOST suitable as the fork choice mechanism capable of providing single slot finality on Ethereum. To achieve this goal, an additional phase, FFG-vote, is added after the fast confirm phase and before the merge, wherein validators cast a vote from the latest justified checkpoint to the target checkpoint. When the optimistic fast confirmation rule is viable, this results in a protocol where blocks are confirmed, immediately become Casper's target checkpoint, and are then voted on to become the latest justified checkpoint. 

Though this solution finalizes a block in every slot, it does not yet finalize a block in the same slot it is proposed. To achieve this without increasing latency to $5\Delta$ rounds per slot, which would be computationally and bandwidth-intensive, a new type of message is introduced -- the acknowledgment~\citep{damato2023simple}. The acknowledgement acts as a message that can be ignored during the actual consensus process, but any participant who desires single slot finality can collect these messages and consider a block finalized once they receive messages from $2/3$ of validators -- a supermajority acknowledgment. 

\paragraph{Streamlining Fast Finality}
\citet{fradamt2023streamlining} creates an additional modification to RLMD-GHOST that sacrifices single slot finality for 3-slot finality with a lower latency protocol and stronger confirmation guarantees. A mechanism for streamlined fast finality with a single proposal and voting phase per slot, instead of having both a fork choice vote and a FFG vote, was introduced. While providing bandwidth and latency improvements, its key contribution is introducing the notion of \textit{strong confirmation}, allowing for the first $\eta$ slots to be reorg resilient, unless $\delta - \frac{1}{2}$ of the stake becomes slashable (say $\delta = \frac{2}{3}$). Additionally, it allows for a proposal at slot $n$ to be justified at slot $n+2$ as long as $n$ and $n+1$ have honest proposers.

In the voting round, validators cast votes for both the tip of the available chain and FFG votes. The source for the FFG vote is the latest justified block, and the target is the latest confirmed block. In this scenario, the latest justified block is strongly confirmed. The view-merge process is also simplified through the use of $\delta$-quorums, which are similar to graded agreements in that they allow validators to have a succinct proof of votes from other validators.

\subsubsection{Remarks}

RLMD-GHOST offers a nuanced approach to vote expiry and asynchrony tolerance, intending to be a suitable replacement for LMD-GHOST in Ethereum's consensus mechanism. With its \(\eta\)-parameterized vote expiration model, RLMD-GHOST is a leading candidate for achieving single slot finality in conjunction with finality gadgets like Casper. One potential area of improvement is its latency. Streamlining the process improves latency, but comes with the tradeoff of 3-slot finality. The primary bottleneck is the view-merge process, which requires additional time in each slot that may not be necessary to achieve similar safety and liveness guarantees. Additionally, without support for subsampling, RLMD-GHOST may require longer round times and novel signature aggregation techniques capable of supporting hundreds of thousands of validators voting in each slot -- a non-trivial task. 


\section{PBFT-Inspired Family}
\subsection{Tendermint}

Tendermint refines classical PBFT consensus protocols for enhanced scalability and liveness in blockchain systems~\citep{latest_buchman_2018}. It is designed for systems with many mutually distrusting nodes communicating over a gossip-based peer-to-peer network. It is the core consensus mechanism in the Cosmos network and Cosmos SDK-based blockchains~\citep{CosmosSDK2021}.

Tendermint modernizes PBFT by relying solely on gossip-based networking rather than point-to-point connections between each node. This system allows the protocol to scale to hundreds if not thousands of validators. Tendermint also introduces a novel termination mechanism to ensure liveness without additional communication overhead.

Tendermint assumes a partially synchronous system model. The protocol tolerates up to $N/3$ Byzantine faults, where $N$ is the number of validators.

\subsubsection{Protocol Phases}

The protocol proceeds in a series of rounds, each with a dedicated proposer selected in a weighted round-robin fashion based on validator voting power. There are three steps per round:

\begin{enumerate}
    \item \textbf{Proposal Phase:} The proposer selects a block and broadcasts a proposal message containing the entire block at the start of the round.
    \item \textbf{Prevote Phase:} Validators broadcast prevote messages for the proposed block hash if they find the proposal acceptable. This vote is sent to all validators through gossip. Otherwise, validators prevote a nil value.
    \item \textbf{Precommit Phase:} If a validator receives \(2N/3\) prevotes for the proposal, it broadcasts a precommit message for the proposed block. Again, this is diffused to all nodes through gossip.
\end{enumerate}

A validator keeps track of several local variables, including its current round, locked value, and valid value -- the latest proposed block that received \(2N/3\) prevotes. If the proposer is honest and the network synchronous, a block can be committed in a single round of the above three steps.

\textbf{Commit Rule:} A validator commits a block if it receives \(2N/3\) precommits for that block in a round.

Otherwise, timeouts trigger moving to the next round to elect a new proposer. The timeout mechanism ensures liveness.

\subsubsection{Features and Observations}
\paragraph{Gossip Network}

Nodes participate in a peer-to-peer gossip network rather than relying on direct connections~\citep{10.1145/41840.41841}. This network is modeled via a "gossip communication" property, which states that any message sent or received by an honest node will be received by all other honest nodes within the \(\Delta\) delay bound after GST.

\paragraph{Finality}

Tendermint's novel termination mechanism, responsible for completing the consensus process for a particular block, ensures liveness without additional communication overhead. This mechanism is accomplished by tracking the latest valid value. The valid value and its associated round number are updated whenever a node observes such a value, and this information is attached to proposals.

After GST, the gossip protocol ensures that locked or valid values propagate to all honest nodes before the round ends. This gossip enables proposers to eventually propose an acceptable block that all nodes will prevote for, ensuring termination/liveness. This mechanism requires no additional messages beyond those in PBFT's standard case, providing termination guarantees with minimal overhead.

A key property of the Tendermint consensus algorithm is instant finality - forks are never created as long as less than one-third of validators are malicious or faulty. When a validator commits a block after receiving \(2N/3\) precommit votes in a round, that block is considered finalized instantly.

Tendermint's instant finality results from a tradeoff - set validator sizes and no dynamic participation support \citep{9833578}. The protocol requires a known, fixed set of validators, and there is no concept of dynamically joining or leaving consensus rounds. It is considered Byzantine behavior if a validator misses proposing or voting in a round. Implementations of Tendermint, such as in the Cosmos SDK, define how the validator set is updated at the application level \citep{9833578}.

Additionally, if more than 1/3 of validators are offline or malicious, the network may halt, and liveness cannot be guaranteed. Compared to LMD-GHOST or other Nakamoto Consensus-based mechanisms, Tendermint prioritizes instant finality and safety over liveness. 

\subsubsection{Remarks}
Tendermint advances PBFT for scalability across a partially synchronous gossip network. The novel termination approach minimizes overhead while providing eventual liveness when less than 1/3 of the validator set is malicious or offline. Though it achieves instant finality, the potential for liveness halts and the lack of dynamic participation are less-than-ideal tradeoffs for a single slot finality mechanism on Ethereum.


\subsection{HotStuff and HotStuff-2}

Introduced by \citet{hotstuff_yin_2019}, HotStuff is a leader-based BFT consensus protocol that improves upon earlier PBFT models, including Tendermint. It operates in the partially synchronous network model, tolerating up to 1/3 Byzantine faults from nodes.

HotStuff is characterized by a three-phase commit structure within a view, optimistic responsiveness, and a linear view change mechanism. The Prepare-Precommit-Commit structure enables linear leader replacement while retaining efficiency during regular operation. Optimistic responsiveness allows leaders to drive consensus at the speed of actual network delay rather than waiting for known upper bounds on delay. The linear view change means that replacing leaders requires only O(n) messages versus O(\(n^3\)) for protocols like PBFT.

\subsubsection{Protocol Phases}

The HotStuff protocol proceeds in a sequence of numbered views, each with a designated leader. The leader aims to extend the blockchain by proposing new blocks and driving consensus on them within the view. To commit a block, the leader steps through three voting phases:

\begin{enumerate}
    \item \textbf{Prepare Phase:} The leader selects the highest quorum certificate it knows, representing a previous round of consensus that satisfied the quorum requirement, and extends it with a new proposal. This is broadcast in a prepare message to all validators.
    \item \textbf{Precommit Phase:} Upon receiving the prepare message and verifying its validity, validators respond with a vote for the proposed block. If the leader receives $2f+1$ prepare votes, where f is the number of faulty or malicious nodes, i.e., votes from 2/3 of validators, it forms a PrepareQC, a quorum certificate that proves the protocol's acceptance. The leader then enters the Precommit Phase and broadcasts the PrepareQC to validators, who will respond with precommit votes.
    \item \textbf{Commit Phase:} If the leader gets $2f+1$ precommit votes, it forms a PrecommitQC and enters this phase. It broadcasts the PrecommitQC and collects commit votes. Once the leader has $2f+1$ commit votes, it forms a final CommitQC that commits the proposal and broadcasts it in a decide message. Validators that receive a decide message consider the proposal in CommitQC as final.
\end{enumerate}

The three communication phases allow HotStuff to safely change leaders with only linear message complexity, a significant advantage over PBFT. The commit rule provides safety based on previous quorum certificates tracking progress.

\subsubsection{Features and Observations}
Compared to protocols like PBFT and Tendermint, HotStuff simultaneously achieves responsiveness and linear view change complexity. PBFT has responsiveness but quadratic view change, while Tendermint is linear but not responsive.

\paragraph{Pacemaker for Liveness}

The HotStuff protocol's safety is decoupled from liveness considerations. Safety follows from the voting rules and commit protocol outlined above. However, a separate component called a pacemaker is required to provide liveness guarantees. The pacemaker is responsible for view synchronization - bringing honest validators to the leader's view so progress can occur.

\paragraph{Enabling Responsiveness}

The critical technique HotStuff uses to retain responsiveness with linear view change is adding a third phase of votes before committing. This consideration provides an extra step where information about the highest locked value propagates to other replicas.

The key difference compared to HotStuff is that Tendermint rounds are coordinated by timeouts rather than Quorum Certificates. Thus, there exists a need to wait for O($\Delta$) time for every block. Tendermint also uses two phases of communication and voting compared to HotStuff's three.

\paragraph{Optimizations in HotStuff-2}
The primary optimization introduced in HotStuff-2 is reducing the number of phases required to commit a block from three down to two~\citep{cryptoeprint:2023/397}. This improvement is achieved by reusing votes across heights in a "chained" structure.

Specifically, HotStuff-2 has the following phased commit structure:

\begin{enumerate}
    \item \textbf{Prepare Phase:} The leader proposes a new block $B$ extending the highest QC it knows. This prepare message is sent to replicas.
    \item \textbf{Precommit Phase:} If replicas find the proposal safe, they respond to the following view's leader with precommit votes for block $B$.
\end{enumerate}

Once the following view's leader assembles $2f+1$ precommit votes for block $B$, it forms a CommitQC that commits block $B$.

Since precommit votes are sent to the leader of the following view instead of the current view, they can serve two purposes:
\begin{itemize}
    \item Precommit block $B$
    \item Prepare block $B+1$ in the next view
\end{itemize}

This pipelining reduces average case commit latency by $1/3$ compared to the original HotStuff protocol. In the optimal case of an honest leader, commitment requires just two communication steps. The two-phase approach does not impact HotStuff's view change procedure. The linear complexity of leader replacement is retained. However, for a stable leader, the steady-state latency is reduced.

\subsubsection{Remarks}

HotStuff is seen as the "golden standard"~\citep{malkhi2023lessons} thanks to its optimal communication complexity, instant finality with no reorganizations, simple design, and foundational role in many in-production mechanisms like DiemBFT~\citep{diemconsensus2021}, Jolteon and Ditto~\citep{gelashvili2023jolteon}, Flow~\citep{hentschel2020flow}, and Narwhal-HotStuff~\citep{danezis2022narwhal}. However, their fixed validator set and lack of asynchrony tolerance, tradeoffs of many PBFT-based mechanisms, pose challenges for implementation as a single slot finality mechanism on Ethereum, as dynamic participation is a unique feature contributing greatly to overall decentralization.


\subsection{Algorand and BA*}

Algorand is a permissionless, proof-of-stake blockchain protocol that utilizes the novel BA* (Byzantine Agreement) consensus mechanism to achieve fast finality, high transaction throughput, and minimal forking~\citep{algorand_gilad_2017}. It aims to address the performance limitations of previous blockchain consensus models like proof-of-work and Classical BFT.

Algorand organizes time into asynchronous rounds, where users agree on a new block to extend the chain in each round while preventing forks. A randomly selected committee of users is chosen to run the BA* protocol, which operates sequentially, coordinating the committee to irrevocably commit to a new block in constant expected time. The sortition process, based on verifiable random functions (VRFs), is crucial for randomness, fairness, and unpredictability in the committee composition. These random committees are critical for providing attack resistance, as adversaries are unable to target participants in advance and their attack would be useless post-committee participation.

Simulations have demonstrated that Algorand can achieve one-minute finality with a throughput significantly higher than Bitcoin, demonstrating minimal performance impact from scaling and a robust tolerance to malicious attacks as long as greater than 2/3 of validators are honest~\citep{10.1145/3132747.3132757}.

\subsubsection{Protocol Phases}

The BA* protocol progresses through a series of timed steps, synchronized across users:

\begin{enumerate}
    \item \textbf{Sortition}: Users are randomly selected for the round's committee via a cryptographic sortition mechanism.
    \item \textbf{Proposal}: A subset of users propose blocks to be added to the blockchain.
    \item \textbf{Reduction}: The committee runs a two-phase reduction to converge on a single block option.
    \item \textbf{Binary Agreement}: A binary voting procedure is executed until a consensus is reached.
    \item \textbf{Finality Check}: The block is considered finalized if a supermajority is reached in the first binary step. Otherwise, it is tentatively confirmed, and later rounds will build on this block and, with overwhelming probability, finalize this round.
\end{enumerate}

Blocks are propagated through a gossip protocol.
\subsubsection{Features and Observations}

\paragraph{Handling Asynchrony and Dynamic Participation}

Algorand safely makes progress even under intermittent asynchrony using techniques such as step timeouts and a recovery protocol. Users are weighted by their stake, supporting open participation while maintaining security. Committee selection is limited to users active in recent blocks, which helps mitigate potential security risks from dormant accounts.

\paragraph{Low Latency Finality}

In the common case, BA* finalizes blocks in just a single voting step. Under synchrony assumptions, safety and finality are reached in a small constant expected number of steps.

\paragraph{Lack of Slashing Penalty}
Unlike other proof of stake protocols, Algorand does not have slashing penalties for maliciously-behaving block proposers or participants. Instead, if a malicious event is detected, the assumption is made that the honest supermajority will not support these actions, and with high probability the following round will have an honest randomly-selected committee to continue building the chain.

\subsubsection{Remarks}

Algorand's BA* consensus design offers fast block finality, fork resistance, high throughput, and resilience to targeted attacks and periods of asynchrony. Since it operates as a complete protocol, not just a consensus layer, adapting its design to fit Ethereum's architecture and addressing potential liveness halts could present challenges. Additionally, even though probability supports the decision to not have a slashing penalty, the addition of one may be beneficial in providing further incentives against acting maliciously in the network.


\section{Total Order Broadcast/Graded Agreement Family}

\subsection{Momose-Ren}

The Momose-Ren (MR) protocol is a recent attempt at solving the open problem of achieving both optimal Byzantine fault tolerance and low latency in a model that allows for the dynamic participation of nodes~\citep{constant_momose_2022}. A pivotal novelty of MR is its introduction of the time-shifted quorum technique to make quorum certificates transferable across nodes with different perceived participation levels. This technique and restoring the transferability of quorum certificates are the core technical contributions of the MR protocol.

In MR's GA protocol, each node starts with an input value. At the end, every node outputs values, each tagged with a grade of 0 or 1. The critical properties satisfied are:

\begin{itemize}
    \item \textbf{Graded consistency:} If an honest node outputs a value \(v\) with grade 1, all honest nodes output \(v\) with at least grade 0.
    \item \textbf{Integrity:} If an honest node outputs a value \(v\), some honest node must have input \(v\).
    \item \textbf{Validity:} All honest nodes output the highest common input value with grade 1.
\end{itemize}

The grade of an output value essentially signifies how many nodes have agreed on it so far. Grade 1 outputs reflect global agreement.

\subsubsection{Protocol Phases}

MR follows the classic view-by-view paradigm adopted by prior BFT protocols. The protocol progresses through iterations called views, each of a fixed duration. The key steps within each view are:

\begin{enumerate}
    \item \textbf{Leader election:} Nodes run a verifiable random function lottery to probabilistically elect a leader for the view.
    \item \textbf{Propose:} The elected leader proposes a new block extending the current candidate block.
    \item \textbf{Graded agreements:} Nodes run five sequential graded agreement instances on the proposed block:
    \begin{itemize}
        \item GA1 and GA2 are used to notarize the block to confirm its uniqueness in the view. These GAs prevent conflicting blocks from being notarized in the same view.
        \item GA3 sets the candidate block for the next view based on its output.
        \item GA4 sets the lock for the next view based on its output.
        \item GA5 finally decides upon the proposed block if it has an output grade of 1.
    \end{itemize}
\end{enumerate}

\subsubsection{Features and Observations}
\paragraph{Time-Shifted Quorum Certificates}

In traditional quorum-based protocols for static networks, a quorum certificate (e.g., a threshold number of votes) certified by one node would be recognized by all other nodes as a valid certificate.

However, this certificate transferability breaks down in the dynamic sleepy model where different nodes can have divergent views of the current participation level at any time. This is because an adversary can selectively announce to subsets of nodes to manipulate their perceived participation.

To address this challenge, MR introduces the concept of time-shifted quorums. The fundamental insight is that while nodes may differ in their current local view of participation, honest nodes can agree on an absolute notion of participation at any past time.

Concretely, the time-shifted quorum certificate construction has nodes broadcast awake messages at two separate times in a 4-round algorithm:
\begin{enumerate}
    \item In Round 1, nodes send "awake" and "echo" messages with a value.
    \item In Round 2, nodes tally the number of echo messages received for each value \(b\).
    \item In Round 3, nodes send another awake message, create another tally of echoes up to that point, and tally \(M_1\) for the number of awake broadcasts they have received
    \begin{itemize}
        \item Let \(E(b)\) be the number of entities from which a party has received a conflict-free echo message for \(b\). Let \(m_1\) be the number of entities from which a party has received awake messages in round 1. If \(E(b) > m_1/2\), nodes vote for that value.
    \end{itemize}
    \item After Round 3, nodes re-tally \(M_1\) and make grade 1 decisions based on the \(E(b)\) from round 2. If \(E(b)\) is still above the majority of awake nodes as determined by \(M_1\), a validator knows that all honest parties awake in round 3 voted for \(b\).
    \begin{itemize}
        \item They also tally the number of votes for \(b\) and \(M_3\), the number of nodes from which they have received the third-round awake message.
        \item If \(V(B) > M_3/2\), where $V(B)$ be the number of entities from which a party has received votes for $b$, nodes output the graded agreement for \(b\) with grade 0.
    \end{itemize}
\end{enumerate}

Through this technique, MR can create quorum certificates that all nodes will recognize despite differences in perceived participation. This missing piece enabled the translation of classic quorum-based techniques to the dynamic sleepy model.

\subsubsection{Remarks}
See section 5.4.2.3 for remarks on both MR and its successor, Malkhi-Momose-Ren.


\subsection{Malkhi-Momose-Ren}
The Malkhi-Momose-Ren (MMR) protocol builds on the Momose-Ren breakthrough to further improve efficiency, security guarantees, and practicality~\citep{byzantine_malkhi_2022}.

\subsubsection{Protocol Phases}
Like MR, MMR operates in repeated views with graded agreements. The key differences within each view are:
\begin{enumerate}
    \item \textbf{Graded Proposal Election (GPE):} A single round election that replaces MR's leader election + propose steps.
    \item \textbf{Two Graded Agreements:} GA' and GA run sequentially rather than requiring 5 GAs like in MR.
    \item \textbf{Early decide:} If GPE outputs grade 1, immediately decide that block after just GPE (in 4\( \Delta \) time).
\end{enumerate}

\subsubsection{Features and Observations}

\paragraph{Enhancements Over Momose-Ren}
The key advantages of MMR over MR include:
\begin{itemize}
    \item \textbf{Lower latency:} MMR reduces the best-case latency from 16\( \Delta \) in MR down to just 4\( \Delta \), comparable to classic non-sleepy protocols, by optimizing the per-view construction.
    \item \textbf{No eventual stability:} MMR eliminates MR's need for an eventual stable participation assumption for liveness through a novel "median of tallies" technique during voting. This technique modifies the voting in GA to remove the need for eventual stable participation by estimating the echo count for each value.
    \item \textbf{Growing corrupt nodes:} MMR allows the number of corrupt nodes to grow proportionally to the overall participation over time, unlike MR, which could only cap corrupt nodes to the minimum participation level.
    \item \textbf{Shorter reviewed history:} Only messages from the immediate last round impact the current round, streamlining the decision-making process.
    \item \textbf{Efficient recovery:} Nodes can recover by fetching messages only from the last few views rather than the unbounded past, facilitating quicker integration into the consensus process.
\end{itemize}

\paragraph{Improved Graded Agreement Protocol}

A key innovation is the modified GA protocol, which provides graded consistency, integrity, and validity like the GA in MR, and includes two additional properties:
\begin{itemize}
    \item \textbf{Uniqueness:} Only one log can be output with grade 1.
    \item \textbf{Bounded divergence:} Nodes output at most two conflicting logs.
\end{itemize}

These new properties are achieved using a \textit{single round} of voting, simplifying the GA process and reducing the latency.

\paragraph{Lower Fault Threshold}

MMR simplifies its GA into a single voting round by weakening the fault tolerance from 1/2 in MR to 1/3, meaning the protocol guarantees safety if less than 1/3 of participants are faulty at any given time. This tradeoff sacrifices security for timing improvements but still achieves a level of security comparable with the previously discussed protocols.

\subsubsection{Remarks on MR and MMR}

Though MR and MMR offer novel consensus approaches through the innovative time-shifted quorum certificate, they are not ideal drop-in replacements for LMD-GHOST in Ethereum due to limitations such as the lack of subsampling support and safety under asynchrony. Further research is needed to explore potential integrations and modifications.


\subsection{D'Amato-Zanolini}

Existing dynamically available Total Order Broadcast (TOB) protocols like MR and MMR enable fluctuating validator participation while retaining safety and liveness using the time-shifted quorum technique. However, MR's multiple voting rounds per decision hamper efficiency, incentivization, and scalability in practice, and MMR sacrifices some fault tolerance for a single voting round. 

\citet{damato2023streamlining} introduce a novel TOB protocol, hereafter referred to as the D'Amato-Zanolini mechanism, that achieves optimal 1/2 adversarial fault tolerance with just a single voting round per decision. This marks valuable progress towards dynamic consensus mechanisms suitable for large-scale, real-world blockchain networks.

\subsubsection{Protocol Phases}
The TOB protocol operates in views spanning 4\( \Delta \) time each. The key innovation is structuring each view \( v \) around a Graded Agreement (GA) instance \( GA_v \) with grades 0, 1, and 2 encapsulating an entire proposal-vote-decision cycle. Candidates, votes, and locks are terms used to reference varying levels of support for blocks.

\textbf{Propose (\( t_v \)):}
\begin{itemize}
    \item Output phase for grade 0 of \( GA_{v-1} \).
    \item Leader proposes extension of \( GA_{v-1} \) grade 0 candidate block.
\end{itemize}

\textbf{Vote (\( t_v + \Delta \)):}
\begin{itemize}
    \item Input phase of \( GA_v \).
    \item Grade 1 output phase of \( GA_{v-1} \).
    \item Logs with grade 1 treated as locks.
    \item Input to \( GA_v \) restricted to lock or block extending lock.
\end{itemize}

\textbf{Decide (\( t_v + 2\Delta \)):}
\begin{itemize}
    \item Output phase for grade 2 of \( GA_{v-1} \).
    \item \( GA_{v-1} \) grade 2 logs are decided.
\end{itemize}

\subsubsection{Features and Observations}
\paragraph{Novel GA Primitive}
A novel GA primitive is introduced satisfying consistency, graded delivery, validity, integrity, and uniqueness while supporting dynamic participation.  

D'Amato and Zanolini join MMR in modifying the GA protocol of Momose and Ren to satisfy the uniqueness property and appropriately treat equivocations. This characteristic is crucial to streamlining the overlaying TOB protocol. Since decided logs are guaranteed to extend locked logs that extend candidate logs, the TOB can encapsulate an entire proposal-vote-decision cycle within one GA instance per view.

\paragraph{GA Algorithm}

The full GA algorithm driving each view is as follows, where:
\begin{itemize}
    \item \( V_\Lambda \) = Logs extending \( \Lambda \)
    \item \( S \) = Input senders
    \item Time thresholds enforce graded delivery
\end{itemize}

\textbf{Input (\( t=0 \)):} Broadcast input log \( \Lambda \)

\textbf{Store (\( t=\Delta \)):} Store all received logs in \( V_\Delta \)

\textbf{Store (\( t=2\Delta \)):} Store all received logs in \( V_{2\Delta} \)

\textbf{Output Grade 0 (\( t=3\Delta \)):}  
If \( |V_\Lambda| > |S|/2 \), output \( (\Lambda, 0) \)

\textbf{Output Grade 1 (\( t=4\Delta \)):} If \( |V_{2\Delta\Lambda} \cap V_\Lambda| > |S|/2 \), output \( (\Lambda, 1) \)

\textbf{Output Grade 2 (\( t=5\Delta \)):} If \( |V_{\Delta\Lambda} \cap V_\Lambda| > |S|/2 \), output \( (\Lambda, 2) \)

The time-shifted quorum technique is essential for supporting dynamic participation. By aligning output thresholds to participation levels across grade transitions, honesty is ensured under fluctuating involvement. D'Amato and Zanolini uniquely apply time-shifted quorums twice in a nested manner. An outer application from \( t=\Delta \) to \( t=5\Delta \) guarantees graded delivery between grades 1 and 2, while an inner application from \( t=2\Delta \) to \( t=4\Delta \) does so for grades 0 and 1. This nested double use of time-shifted quorums is pivotal to reducing the number of voting rounds per decision.

\paragraph{Single Voting Round}
The single voting round per decision enhances efficiency, incentives, and scalability versus previous protocols requiring multiple rounds. With expected uniform votes per round, vote aggregation is also better supported in large networks.

There is a 2\( \Delta \) stability requirement for validators, which aligns with practical state recovery limitations upon rejoining while upholding security. This reasonable tradeoff is warranted for such substantial efficiency gains.

\paragraph{Asynchrony Resilience}

In their paper, D'Amato and Zanolini showcase how their TOB protocol can be augmented to withstand bounded periods of asynchrony using techniques adapted from their prior work on RLMD-GHOST~\citep{damato2023recent}.

This is achieved by adding an expiration period \( \eta \) for votes and modifying the GA to utilize latest messages from the previous \( \eta \) instances in determining outputs. The paper's appendix provides proof that this does not break this mechanism's other properties.

\subsubsection{Remarks}
By streamlining dynamic TOB, D'Amato and Zanolini further the practicality of highly participatory consensus protocols suitable for real-world decentralized blockchains. Their innovations mark valuable progress towards robust and efficient large-scale dynamic consensus mechanisms. With 1/2 tolerance to adversaries and only one voting round, this mechanism is a promising candidate for adoption as an LMD-GHOST replacement in Ethereum.
\chapter{Single Slot and Fast Finality with D'Amato-Zanolini}
\section{RLMD-GHOST SSF/fast finality accomplishments}
As explored in~\citet{damato2023simple} and section 5.2.2.2.4 of this work, RLMD-GHOST is capable of providing the basis for a single slot finality mechanism. This is accomplished through the use of an additional FFG voting round, mimicking the 2-phase voting system seen in traditional instant finality BFT protocols. With a head-vote fast confirmation round followed by an FFG-voting round, a proposal can be justified within its proposal slot, and be considered finalized either through an additional voting round or informally through an acknowledgment gossiped among sub-networks interested in benefitting from SSF. It is also possible to streamline the head vote and FFG vote in the same round, and get better latency at the cost of 3-slot finality ~\citep{fradamt2023streamlining}.

\section{Motivation for further research}
Though these area an elegant solution, room for improvement remains. Namely, the process of view synchronization through view-merge has high bandwidth and latency requirements, and it would be ideal if the validators could come to the same conclusion without having to first synchronize their views. In~\citet{damato2023streamlining}, a new mechanism was proposed that uses time-shifted quorums to create a Graded Agreement-based total order broadcast protocol with $1/2$ resistance to adversarial participants, reorg resilience, a single instance of GA, and a single voting round. This mechanism can also be augmented with a vote expiry period (similar to RLMD-GHOST) to induce resistance to temporary asynchrony, using techniques from ~\citet{damato2023improving}. With this protocol as our basis, we can explore modifications analogous to those for RLMD-GHOST that give it single slot finality or streamlined fast finality, each with its respective trade-offs that must be considered before the implementation of one of these systems can take place.

The rest of this section will be organized as follows:
\begin{itemize}
    \item Step-by-step of GA-based SSF, including adding the fast-confirm and FFG-vote rounds.
    \item Streamlined GA-based consensus with strong confirmations.
\end{itemize}

\section{Constructing GA-based SSF}
\subsection{GA-Based TOB}
Starting with the D'Amato-Zanolini mechanism described in section 5.4.3 as a basis, we can introduce similar concepts as those in~\citet{damato2023simple} to achieve SSF, albeit at the cost of added complexity and latency. First, this section will provide a brief overview of the total order broadcast protocol leveraging graded agreements that is capable of instant decisions. We'll then introduce an SSF-enabled mechanism informally proposed in \citet{damato2024ga}. Finally, we will propose a mechanism for streamlined fast finality. 

This TOB protocol proceeds in slots of four rounds each, for a total of $4\Delta$ time per slot. The GA lasts $5\Delta$, and each slot's GA bleeds into the first round of the following slot.

\subsection{Explicit specification without the GA black box}
\begin{enumerate}
    \item \textbf{Propose ($t=t_s$)}: Leader proposes a block extending the current candidate, meaning the highest block such that it has more than $1/2$ of the votes.
    \item \textbf{Vote ($t=t_s+\Delta$)}: Vote for proposal extending the lock, the highest block that has votes from $t_{s}-\Delta$ AND the current round from more than half of the current perceived participation.
    \item \textbf{Decision ($t=t_s+2\Delta$)}: Decide on the block that has votes from $t_{s}-2\Delta$ AND the current round for more than half of the current perceived participation.
    \begin{enumerate}
        \item Store the current vote count for use in $4\Delta$ rounds from now.
    \end{enumerate}
    \item \textbf{($t=t_s+3\Delta$)}: Store the current vote count for use in $2\Delta$ rounds from now.
\end{enumerate}

\subsection{Adding SSF}
In the previous construction, decisions about a block in slot $s$ are made, in the best-case scenario, in slot $s+1$. For single slot finality, blocks naturally need to be proposed, justified, and finalized within the same slot.

As a part of accomplishing SSF in an efficient manner, it first makes sense to remove the grade 2 GA from the protocols, sacrificing deterministic safety for probabilistic safety. This not only reduces the number of rounds in the GA but allows for the inclusion of a fast confirmation rule that can be included in the same slot. The previous decision rule could be kept, but the only advantage would be support for low participation environments, where finality may not be of greatest concern.

\subsubsection{2-grade Graded Agreement}
\begin{enumerate}
    \item \textbf{($t=0$):} Input block
    \item \textbf{($t=\Delta$):} Store the current voting level as $V$
    \item \textbf{($t=2\Delta$):} Output GA with grade 0 if current voting level minus equivocating votes is greater than half participation
    \item \textbf{($t=3\Delta$):} Output GA with grade 1 if voting level from round 2 minus current equivocating voters is greater than half current participation
\end{enumerate}

Note that in the TOB protocol from section 5.4.3.1, the decision phase is only concerned with the grade 2 output, which is no longer a part of this new GA. Thus, we can remove this phase and create a more succinct TOB protocol only using grades 0 and 1. 

\subsubsection{TOB protocol with probabilistic safety, $3\Delta$ rounds}
\begin{enumerate}
    \item \textbf{Propose ($t=t_s$)}: Leader proposes a block extending the highest block output with grade 0 from the previous GA.
    \item \textbf{Vote ($t=t_s+\Delta$)}: This round's GA starts, and a block that extends the lock (grade 1 output) from the previous round is input to the current GA.
    \item \textbf{($t=t_s+2\Delta$)}: Store current voting level for the GA protocol.
\end{enumerate}

This protocol still provides the guarantee that candidates extend locks. Therefore, honest votes are given to honest proposals. It also provides reorg resilience with probabilistic safety for the $\kappa$-deep confirmation rule.

\subsubsection{Adding Fast-Confirms}
With this probabilistically-safe protocol developed, the next step is to include a fast confirmation rule that will instantly confirm a block whose subtree, in the current slot, has received $2/3$ votes from the whole validator set. With instant confirmation, we will then be able to direct an FFG vote towards a block in the same slot in which it was proposed. 

\begin{itemize}
    \item Between the second round ($\Delta$) of slot $s$ and the second round of slot $s+1$, validators keep track of $b_C$: the highest block with $2/3$ votes from slot $s$.
    \item They also track the corresponding quorum certificate that succinctly proves the $2/3$ votes for the subtree of $b_C$.
    \item Since these blocks have more than the required votes, we must modify the voting rule to acknowledge their candidacy for fast confirmation. This gives us safety for our confirmation rule.
\end{itemize}

\subsubsection{Fast-confirm TOB Protocol}
\begin{itemize}
    \item In \textbf{Propose ($t=t_s$)}, the leader proposes a block, $b'$, that has $b_C$ as its prefix, where $b_C$ has over $1/2$ support in the current view of votes and voters. It also includes $b_C$ and $Q_C$, a valid certificate for $b_C$.
        \begin{itemize}
            \item If there's no block satisfying these properties, then the leader proposes one that simply extends $b_C$.
        \end{itemize}
    \item In \textbf{Vote ($t=t_s + \Delta$)}, if a valid proposal is received and the proposed block $b^p_C$ has $b_C'$, the block that the voter currently sees as most recent, as a prefix, then the voter updates their view of the most recent block, labeled as $b_C'$, to $b^p_C$.
        \begin{itemize}
            \item Let the slot $s$ \textit{lock} $L_s$ be the highest block that has this new $b_C'$ as a prefix and has over half support from the union of votes now and in the previous round of voting, at time $t_s - \Delta$ (equivalently, $t_{s-1} + 3\Delta$).
            \item If no such block exists with the required support to be considered, set $L_s$ to $b_C'$, meaning the lock is the newly-proposed block $b^p_C$.
            \item Vote for a proposal extending $s$'s lock, or the lock itself if no such proposal exists.
        \end{itemize}
    \item \textbf{Fast confirm (t=$t_s + 2\Delta$):} confirm the block proposed and received sufficient support in the current round. For the leader, this is $b'$, and for voters this is $b^p_C$ they received and accepted as $b_C'$. 
    \item $t_s + 3\Delta$: Perform actions required by the GA.
        \begin{itemize}
            \item Store current voter support $V$ as $V'$.
            \item Set $b_C' = b_C$.
        \end{itemize}
\end{itemize}

\subsubsection{Adding FFG-Vote}
Now, there is a protocol that's able to confirm blocks in the slot in which they're proposed with a $2/3$ majority. If we add an FFG-vote phase, this will be sufficient to follow a similar model to RLMD-GHOST and have FFG votes be cast with the current slot as the target slot.

To do so, we add an FFG-vote phase directly after the fast-confirm. In it, the validator casts a vote using the latest justified block, with the target being the highest confirmed descendant of the latest justified block. In the optimistic case, the latest justified block should be from slot n-1, and the highest confirmed descendent from the current slot n. In our proposals and votes, we also include checkpoints, which are pairs of blocks and justification slots ($b, s$). Validators store the set of justified checkpoints $J$, specifically taking note of the latest justified checkpoint $LJ$. We also modify the definition of $b_C$ to be the highest block that has $b_{LJ}$ as a prefix with a quorum certificate from slot $s$, if there is one, and $b_C = b_{LJ}$ otherwise. Proposals now also include ${LJ}$. 

\subsubsection{TOB with FFG Vote (SSF-capable)}
\begin{itemize}
    \item \textbf{Propose ($t=t_s$)}: Leader $p$ proposes a block $b'$ that has $b_C$ as its prefix with over half support in the current view of votes and voters, also including $b_C, Q_C, {LJ}$ in the proposal.
        \begin{itemize}
            \item If no block satisfies these properties, then the leader proposes one that simply extends $b_C$.
        \end{itemize}
    \item \textbf{Vote ($t=t_s + \Delta$)}: if a valid proposal is received:
        \begin{itemize}
            \item First, if the proposed block's latest justified checkpoint ${LJ}^p$ is in the voter's set of justified checkpoints, and its slot is more recent than their current latest justified ${LJ}'$, update ${LJ}'$ to ${LJ}^p$.
            \item If $b_C'$ is a prefix of the proposed $b_C^p$, set $b_C' = b_C^p$.
            \item Let the slot $s$ lock $L_s$ be the highest block that has this new $b_C'$ as a prefix and has over half support from the union of votes now and at time $t_s - \Delta$ ($t_{s-1} + 3\Delta$).
                \begin{itemize}
                    \item If no such block exists with the necessary support to be $L_s$, then set $L_s$ to the current $b_C'$, which now is the newly updated block $b_C^p$.
                \end{itemize}
            \item Vote for a proposal extending $L_s$, or the lock itself if no such proposal exists.
        \end{itemize}
    \item \textbf{Fast confirm and FFG-vote (t=$t_s + 2\Delta$)}: confirm the block that was proposed in this slot, $b_C$. Cast FFG vote from source ${LJ}' \to (b_{C,s})$.
    \item \textbf{($t_s + 3\Delta$)}: Perform actions required by the GA.
        \begin{itemize}
            \item Store the latest justified checkpoint as ${LJ}'$.
            \item Store current voter support $V$ as $V'$.
            \item Set $b_C' = b_C$.
        \end{itemize}
\end{itemize}

Under optimistic conditions, this protocol justifies a block within the slot in which it is proposed. This provides for 2-slot finality, but single-slot remains elusive. As a solution, consider the adoption of the acknowledgment message from \citet{damato2023simple}. After the FFG-Vote, a validator broadcasts an acknowledgment if the target was justified in the current slot. Then, add a slashing condition that says a validator will not do an FFG vote for a slot less than the most recent where it voted. These acknowledges provide a form of opt-in single slot finality without the overhead of an additional round.

\section{Streamlined Fast Finality}
Instead of adopting single slot finality, one alternative would be accept multi-slot finality with lower latency, which may provide sufficient economic finality guarantees while also reducing latency by streamlining the consensus procedure. We can remove one round by merging the head vote round with the FFG vote, similar to the approach taken in \citet{fradamt2023streamlining} for RLMD-GHOST. 

A simplified explanation of how the protocol would operate is as follows:
\begin{itemize}
    \item \textbf{Propose ($t=t_s$)}: Leader $p$ proposes a block $b'$ that has $b_C$ as its prefix with over half support in the current view of votes and voters, also including $b_C, Q_C, {LJ}$ in the proposal.
        \begin{itemize}
            \item If no block satisfies these properties, then the leader proposes one that simply extends $b_C$.
        \end{itemize}
    \item \textbf{Vote ($t=t_s + \Delta$)}: if a valid proposal is received:
        \begin{itemize}
            \item First, if the proposed block's latest justified checkpoint ${LJ}^p$ is in the voter's set of justified checkpoints, and its slot is more recent than their current latest justified ${LJ}'$, update ${LJ}'$ to ${LJ}^p$.
            \item If $b_C'$ is a prefix of the proposed $b_C^p$, set $b_C' = b_C^p$.
            \item Let the slot $s$ lock $L_s$ be the highest block that has this new $b_C'$ as a prefix and has over half support from the union of votes now and at time $t_s - \Delta$ ($t_{s-1} + 3\Delta$).
                \begin{itemize}
                    \item If no such block exists with the necessary support to be $L_s$, then set $L_s$ to the current $b_C'$, which now is the newly updated block $b_C^p$.
                \end{itemize}
            \item Vote for a proposal extending $L_s$, or the lock itself if no such proposal exists.
            \item Cast FFG vote from source ${LJ}' \to (b_{C,s})$, the latest confirmed block.
        \end{itemize}
    \item \textbf{($t_s + 2\Delta$)}: Perform actions required by the GA.
        \begin{itemize}
            \item Store the latest justified checkpoint as ${LJ}'$.
            \item Store current voter support $V$ as $V'$.
            \item Set $b_C' = b_C$.
        \end{itemize}
\end{itemize}

In this protocol, if slots $s$ and $s+1$ have honest proposers, in the best case, a block proposed in slot s will be the latest confirmed in slot $s+1$, and will be the latest justified in $s+2$. This approach provides finality three slots after a block is proposed, which is still significantly better than Gasper's current 64-slot minimum and saves an additional $\Delta$ in time per slot.
\chapter{Cumulative Finality}
Thus far, finality has been considered as a binary option - either a block is finalized or it is not. However, finality can have different levels and build up over time. This concept is known as cumulative finality.

The two most likely candidates for replacing LMD-GHOST, namely RLMD-GHOST and D'Amato-Zanolini, provide a specific type of finality called economic finality. As mentioned in section 4.1, perfect cryptographic finality is the only irreversible type, while economic finality can be undone as long as malicious actors have access to and are willing to burn enough stake. 

In the binary economic finality model, where something is either finalized or it is not, there is no additional penalty or difficulty when attempting to break finality at different levels of depth, e.g., at 64 blocks back or 256 blocks back. That being said, the farther back a block resides, the longer it will take for new blocks to be proposed and built to catch up to the current longest chain and ultimately become the longest -- this $\kappa$-deep confirmation rule provides the basis of security in Nakamoto consensus.

In a single-slot finality protocol using 2/3 of validator signatures per slot, cumulative finality is not necessary, as this system already provides strong guarantees with a supermajority approving each block and consequently requiring 2/3 of staked ETH to be burned to reverse this decision. However, if single slot finality were achieved by sacrificing full sampling, as mentioned in section 4.4, the strength of the economic finality would also suffer. In particular, for a rotating participation approach, early calculations demonstrate that the attack cost would be about 1/32 of the entire staked ETH~\citep{buterin2023sticking}. 

Implementing cumulative finality could increase this number each time a committee is selected, and validators could both attest to the previous finalized blocks and the latest justified block. As more committees with unique ETH vote, the cost of reorging older finalized blocks increases. Once 2/3 of staked ETH attests to a block over several committees, reorging that block would have the same cost as in the current consensus mechanism. 

~\citet{buterin2021cumulative} has previously proposed a cumulative committee-based finality design, but approach 3 in his most recent model, hereafter referred to as the rotating participation approach, has some modifications from this design ~\citep{buterin2023sticking}. Rotating participation's key contribution is creating committees with much more staked ETH, leading to greater economic security per slot. Along with allowing for variable amounts of stake, rotating participation permits validators with more than a certain variable quantity of ETH to participate in the committee during every slot, and validators with less than this quantity have a chance of participating proportional to their stake. In a cumulative economic finality system, a validator participating in a finalization vote for a block more than once, such as if it was in back-to-back slots, would result in a "double-attest" where the same ETH would be twice counted towards a block's economic finality. Say that this validator votes for the finalization of block A in slot n then is on the committee in slot n + 1 and votes for a block building on A. In the cumulative model, if this validator later decides to act maliciously and equivocate in slot n, it should be slashed for twice its value since it has committed to the cumulative finality twice. However, Ethereum slashing penalties grow in magnitude as the number of malicious actors in a given time frame increases, so if a large coordinated slashing event occurs the validator may not have enough at stake to be properly penalized for their actions.

Modifying the rotating participation approach to work with cumulative finality requires more technical complexity but provides stronger security guarantees. Specifically, something must be done to solve the double-attest problem while allowing for variable amounts of staked ETH. The primitive solution to this problem is to ensure that no validator is involved in a cumulative finalization decision more than once and can only become eligible for validation after a block's cumulative economic finality is fully saturated. This is reminiscent of the current Gasper model, wherein each validator is active once and only once in each 32-block epoch.

Three primitive modifications allowing for cumulative finality in approach 3 are proposed:
\begin{enumerate}
    \item Revert the proposal that validators over a certain amount of stake are always included.
    \item For validators with more than $M$ ETH, split their stake among several slots so they are always validating and always contributing to finality.
    \item Keep track of which validators have already contributed to cumulative stake and prevent their vote from counting towards previously-attested blocks while still allowing them to vote in every slot.
\end{enumerate}

Modification 1, while simple, neglects to properly align validator and protocol incentives. The point of staking larger amounts is to be chosen for votes and proposals more often, thus earning greater staking rewards. Compared to the binary economic security approach, each individual validator's stake would likely be lower. This scenario may offer fewer rewards for stakers than the current model since users would be chosen at most for 1 of every 32 blocks. Additional work must be done to ensure that the validators with the highest probability of being chosen would not always be chosen early into the 32-block period and instead properly spaced out to prevent potential coordinated attacks. 

Modification 2 has as a consequence that validators' large stakes, which they are willing to put as collateral towards one block in one slot, are split up, resulting in lower economic security on a per-block basis. Though finality would consistently accumulate every slot, the attack cost would be significantly lower after the first few slots than in the native rotating participation approach, and would take dozens of slots to get to this minimum threshold. Say that a validator is contributing to every slot. Then, their stake must be split over at least as many slots as it takes for a block to become cumulatively finalized. By taking the assumptions from Buterin's initial post, where the top 512 validators with greater than M stake account for 2,359,296 ETH and the rest of the randomly sampled stakes in each slot committee accounting for 262,144 ETH, it would take significantly longer to achieve meaningful cumulative finality than in the other models~\citep{buterin2023sticking}. For example, using 32 slots, the top validators contribute 73,728 ETH per slot and the rest at 262,144 ETH, the result is 10,747,904 ETH staked in support of a block by the end. Though this is quite high, the attack cost after one block is around 100,000 ETH - not strong enough to offer meaningful guarantees. 

Modification 3 would result in a system with the same per-block guarantees as the native rotating participation approach but also leads to a slow increase in cumulative finality, primarily driven by small solo stalkers who are randomly chosen and don't have the stake to contribute consistently. After a block is proposed, confirmed, and finalized, the large stakers involved in every slot and the committee-specific randomly chosen smaller stakers would attest to this block with all of their stake, which would be recorded. At this point, the attack cost would be the same as in the binary economic finality approach with subsampling -- namely 900k ETH. In future rounds, their previous participation towards a block's finality would prevent them from double-attesting. Instead, any new validators in the slot's committee would contribute towards accumulating more finality. Given that the rotating smaller validators are responsible for a minority portion of the stake in any given slot, this mechanism would result in a slower time-to-saturated finality and greater technical complexity to account for which stakers have already attested to a specific slot. The primary question to consider is if the technical complexity would be worth it -- would there be meaningful cumulative finality built up over time from solo stakers alone? Given that the randomly sampled smaller stake would add up to approximately 262,144 ETH per slot per initial calculations in~\citet{buterin2023sticking}, every slot of completely new validators would result in an additional 90,000 ETH in attack cost for a given block. 

There are additional technical constraints that must be considered for these approaches. Particularly, how would a validator's support of a block be tracked and proliferated across the network? The primary reason for reducing the number of validators per slot was to alleviate the burden of signature aggregation, and having validators attest to every block, even a fraction per slot, could reintroduce these problems. Another option might be to modify the finalization vote to state that if a validator votes for any block, they are also cumulatively adding to the support that block's ancestors up to the last one it recorded a finalization vote for. It is left to future work to determine the practical implementation of a system like the one proposed in Modification 3. 

\chapter{Conclusion}
This thesis systematically evaluated consensus protocols and mechanisms for achieving single slot finality (SSF) in Ethereum. Motivated by limitations in Ethereum's current Gasper protocol, which relies on the vulnerable LMD-GHOST fork choice rule and Casper FFG overlay for economic finality, the goal was to assess alternative consensus mechanisms and protocol designs capable of faster finality.

After providing an overview of useful background information, an LMD-GHOST replacement's desired traits were outlined -- namely, drop-in compatibility with Ethereum's architecture, accommodation of validator subsampling, and resistance to temporary asynchrony. A thorough study of prominent consensus protocol families was subsequently undertaken, scrutinizing individual mechanisms based on their liveness guarantees, finality capabilities, and suitability for Ethereum's vision.

This thesis explored several pathways to introducing single slot finality into Ethereum, spanning the protocol and application layers. Leveraging emergent cryptographic primitives like one-shot signatures could enable provable SSF by having validators create unforgeable block attestations. However, quantum computing barriers currently render this technique aspirational. Modifying the core consensus mechanism proves most viable but also time-intensive regarding rigorous testing and coordination. As a bridge, the EigenLayer framework's actively validated services model offers clever workarounds before a fork materializes. Still, challenges around incentive alignment with non-participating validators and protocol rules persist.

Propose-vote-merge protocols like RLMD-GHOST demonstrate promise in replacing LMD-GHOST thanks to carved epochs, fork choice on block weight, and configurable vote expiry periods that balance responsiveness with asynchrony tolerance. However, computational overhead from view synchronization remains a barrier. Among PBFT-inspired designs, the classic tradeoff between dynamic participation and instant finality persists, rendering protocols like Tendermint and HotStuff inapplicable without modification. The single-vote efficiency of D'Amato-Zanolini's TOB mechanism stands out as an appealing foundation for SSF on Ethereum. 

With D'Amato-Zanolini selected as the base consensus mechanism and potential LMD-GHOST replacement, this thesis simulated SSF by incorporating fast confirmation and FFG voting rounds, following the model conceived for RLMD-GHOST. In an alternative approach, latency and bandwidth refinements were achieved via modifications for a streamlined voting structure that retains 3-slot finality. Additionally, a framework was established for constructing cumulative finality atop committee-based approaches to enhance probabilistic guarantees.

In totality, this thesis puts forth actionable insights, recommendations, and parameters for advancing single slot finality objectives on Ethereum. The discourse covers the expanse of existing protocols, identifies salient tradeoffs, and plots a course towards augmentations that preserve decentralization without sacrificing security or efficiency.

\section{Future Work}
With the foundational groundwork laid through this thesis, several promising directions emerge for furthering research on SSF-suitable consensus mechanisms and their ultimate adoption in the Ethereum protocol:

\textbf{Formal specification of streamlined D'Amato-Zanolini}: While the operational logic has been conveyed, formally specifying safety and liveness proofs would reinforce the suitability of the optimizations presented in Chapter 6. Cryptographic formalisms would lend mathematical rigor.

\textbf{Economic modeling and simulations}: Quantiative analysis of factors like staker reward dynamics and attack costs given finality subsampling would enable an informed selection of parameters that create favorable conditions for cumulative finality constructions and elucidate the consequences of allowing different degrees of subsampling.

\textbf{Client implementation and benchmarks}: Realizing variants of the studied protocols on various Ethereum clients and benchmarking performance against Gasper on metrics like transactions per second, latency, bandwidth overhead, and fault tolerance would provide vital data on the feasibility and practical implications of a fork choice rule change.

\singlespacing
\bibliographystyle{apalike}
\bibliography{MyLibrary}
\end{document}